\documentclass[prc,preprint,showpacs,tightenlines,%
                floatfix,amsfonts,nofootinbib]{revtex4}

\usepackage{latexsym}
\usepackage{graphicx}

\begin{document}

%
%
%
%
\def\psibar{\overline\psi}
\def\tr{\mathop{\rm tr}\nolimits}
\def\Tr{\mathop{\rm Tr}\nolimits}
\def\zz{\hphantom{-}}          
\def\xx{\hphantom{0}}          


\title{Loop Corrections and Naturalness in a\\
       Chiral Effective Field Theory}

\author{Jeff McIntire}\email{\texttt{oberonjwm@yahoo.com}}
\author{Ying Hu}\email{\texttt{yihu_us@yahoo.com}}\altaffiliation[permanent
address: ]{Motorola Inc., 1501 W. Shure Dr., Mail Drop:~3--2G, Arlington
Heights, IL 60004.}
\author{Brian D. Serot}\email{\texttt{serot@indiana.edu}}
 \affiliation{Department of Physics and Nuclear Theory Center
             Indiana University, Bloomington, IN\ \ 47405}

%
\author{\null}
\noaffiliation

\date{\today}
%

\begin{abstract}
The loop expansion is applied to
a chiral effective hadronic lagrangian; with the techniques of
Infrared Regularization, it is possible to separate out the
short-range contributions and to write them as local products of
fields that are already present in our lagrangian. (The appropriate
field variables must be re-defined at each order in loops.) The
corresponding parameters implicitly include short-range effects to
all orders in the interaction, so these effects need not be
calculated explicitly. The remaining (long-range) contributions that
must be calculated are nonlocal and resemble those in conventional
nuclear-structure calculations.
Nonlinear isoscalar scalar $(\sigma)$ and vector $(\omega)$ meson
interactions are included, which incorporate many-nucleon forces
and nucleon substructure.
Calculations are carried out at the two-loop level to illustrate
these techniques at finite nuclear densities
and to verify that the coupling parameters remain natural when
fitted to the empirical properties of equilibrium nuclear matter.
Contributions from the $\omega N$ tensor coupling are also discussed.

\end{abstract}

\smallskip
\pacs{24.10.Cn; 21.65.+f; 24.10.Jv; 12.39.Fe }

\maketitle

\section{Introduction}

In recent years, density functional theory (DFT) has emerged as a
powerful tool for describing the structure of nuclei. Based on the
seminal works of Kohn, Hohenberg, and Sham
\cite{ref:Ho64,ref:Ko65,ref:Ko99}, DFT was originally developed for
use in condensed matter physics; interestingly, it was discovered
that it also provides a systematic framework for modeling the
low-energy sector of the strong interaction
\cite{ref:Dr90,ref:Sp92,ref:Sc95,ref:Fu97,ref:Fu98,ref:Se97,ref:Wa04}.
Central to DFT is the notion that, if the energy functional is
considered a functional of the density, then by minimizing the
energy one can acquire the exact ground-state density
\cite{ref:Ho64,ref:Ko99}. Furthermore, if the quantum many-body
equations are replaced by single-particle equations with local,
classical fields, one can exactly reproduce the total energy, scalar
and vector densities, and chemical potential
\cite{ref:Ko65,ref:Al85}. Thus the many-body problem is reduced to
finding the proper form of the energy functional. As it turns out,
the lagrangian of quantum hadrodynamics determines just such a
functional.

Quantum hadrodynamics (QHD) is a low-energy effective theory of the
strong interaction that uses the observed degrees of freedom at this
energy scale (hadrons). It has evolved from the early model of
Walecka \cite{ref:Wa74} into more modern incarnations that are based
on chiral effective field theory (EFT) and DFT
\cite{ref:Fu97,ref:Fu98,ref:Se97,ref:Fu99,ref:Fu03,ref:Se04,ref:Fu04,ref:Wa04}.
The QHD energy functional is constructed as an expansion in powers
of the mean meson fields, which are in fact Kohn--Sham potentials,
over a heavy mass, such as the chiral symmetry breaking scale,  and
includes all possible terms consistent with the underlying
symmetries of QCD. Field redefinitions are used to move the
complexity of the problem (e.g., many-body forces and nucleon
substructure) into the meson field self-interactions. As this is an
effective theory, the terms are characterized by unknown
coefficients. Once all the dimensional and combinatorial factors
have been removed, however, the remaining dimensionless coefficients
are assumed to be \textit{natural}, or of order unity
\cite{ref:Ma84,ref:Ge93,ref:Fr96,ref:Fr97,ref:Fu97,ref:Fu98}. The long-range dynamics
are included explicitly, and short-range physics is contained in the
parametrization; therefore, this theory incorporates the natural
separation of length scales. While QHD in principle contains all
possible terms consistent with the underlying symmetries of the
system, in practice it is a perturbative expansion for the energy
functional that can be truncated at a manageable level (since the
ratio of the mean fields to the heavy mass is small). The now finite
number of coefficients are fixed by experimental data, and this
theory can then be used for predictive purposes
\cite{ref:Mc02,ref:Mc04,ref:Mc05,ref:He02,ref:He03,ref:He03a,ref:He04}.

QHD is a strong-coupling theory. Unlike QED, there is no obvious
asymptotic expansion to use to obtain results and refine them
systematically. It is thus unknown whether QHD permits any expansion
for systematic computation and refinement of theoretical results.
One possibility is the loop expansion, which was partially explored
in \cite{ref:Fu89,ref:We90,ref:Ta95,ref:Mc07}.

Why a loop expansion? The loop expansion is a simple and well-developed expansion scheme in
powers of $\hbar$ that is derived from the path integral
\cite{ref:Di33,ref:It80,ref:Se86,ref:Fu89,ref:Co73,ref:Il75,ref:Co77}.
The mean meson fields are included non-perturbatively,
and the correlations are included perturbatively.
Therefore, one can analyze the many-body effects order by order.
Previous work in QHD has shown that the mean-field terms dominate the nuclear energy, and exchange
and correlation effects do not significantly modify the energy or nucleon self-energies, at least
for states in the Fermi sea \cite{ref:Se86}.
We stress that we are not certain that the loop expansion is
practical; the answer to this question is left for future consideration.
However, the loop expansion has the advantage that it is fairly easy to separate the
short-range and long-range dynamics and to analyze their structures.

The loop expansion was investigated in the simplest case at the two-loop level
\cite{ref:Fu89,ref:Mc07} (only the most rudimentary
couplings were retained).
This was a severe truncation of the EFT lagrangian and implies that we were interested only in
the lowest level of accuracy.
Here we are also concerned with the loop expansion only up to two-loop order
(consideration of higher-loop effects is left for the future),
but now nonlinear meson self-couplings and tensor terms are retained.
Note that the nonlinear terms incorporate short-range many-body nuclear forces, vacuum
dynamics, and the effects of nucleon substructure.
The extension of the level of truncation to include these terms allows us to describe nuclear
matter (and presumably nuclei) with greater accuracy.

The result of the loop expansion up to the two-loop level is the mean field energy density
\cite{ref:Fu97,ref:Fu98,ref:Se97} plus a series of integrals.
We can categorize these integrals in two groups: baryon loops and purely
mesonic loops.
The baryon loops are integrals at the two-loop level that contain two factors
of the baryon propagator and one factor of a meson propagator.
Each nucleon propagator can be separated into a Feynman and a
Density part \cite{ref:Se86,ref:Fu89}.
This is accomplished by incorporating the proper pole structure of the propagator.
The Feynman part describes the propagation of a baryon or an antibaryon; the Density part involves
only on-shell propagation in the Fermi sea and accounts for the exclusion principle.

As before \cite{ref:Mc07}, the two-loop baryon integrals can each be separated into three
distinct parts: the exchange, Lamb-shift, and vacuum-fluctuation contributions.
The exchange term has two factors of the Density portion of the nucleon propagator.
The Lamb-shift term contains both Feynman and Density parts of the nucleon propagator.
The vacuum-fluctuation term contains two factors of the Feynman propagator.
The Lamb-shift and vacuum-fluctuation parts are short-range physics,
and it was shown in the simple case of \cite{ref:Mc07}
that they can be expressed as a sum of terms that are already present in the
EFT lagrangian.
This is an effective theory, so the coefficients of these terms are not yet determined;
thus, the two-loop terms are just absorbed into terms already present in the lagrangian
and are not to be calculated explicitly \cite{ref:Mc07}.
As a result, only the exchange portion of the two-loop integrals must be calculated explicitly.

In addition, a number of purely mesonic loops arise when nonlinear meson field self-interactions
are included.
These terms, however, can be expressed as a power series in the meson fields with undetermined
coefficients.
As before, these terms are just absorbed into the lagrangian parameters
and should not be calculated explicitly.

It is interesting to note that the natural separation of the length
scales in QHD and the parametrization of the short-distance dynamics
is similar to Infrared Regularization
\cite{ref:Be99,ref:Be00,ref:Sc03,ref:El98,ref:Ta96,ref:Ge03,ref:Sc04,ref:Ge04,ref:Sc06,ref:Dj06}. In Infrared
Regularization, loop integrals are separated into infrared regular and
singular contributions (corresponding to short- and long-range
physics). The regular portion is expanded as a power series in the
momentum and is absorbed into the unknown coefficients of the
underlying lagrangian; the singular part must be calculated
explicitly. This is directly analogous to what was done for the loop
expansion in QHD \cite{ref:Mc07}.

One question that remains unanswered is how the addition of these
many-body correlations to the energy density affects the naturalness
assumption, which states that once all the dimensional and
combinatorial factors in a given term are accounted for, what
remains is a dimensionless coefficient of order unity
\cite{ref:Ma84,ref:Ge93,ref:Fr96,ref:Fr97,ref:Fu97,ref:Fu98}. Once one decides on a
level of truncation in the underlying lagrangian, the now finite
number of unknown constants can be parametrized by the data. Thus,
one purpose of this work is to prove that sets of parameters exist
at different levels of truncation in the two-loop energy density in
which the parameters are all natural and thereby show
that the naturalness assumption holds. For the purposes of this
work, we will consider only uniform, symmetric nuclear matter, as in
\cite{ref:Mc07}. Here, however, we also include nonlinear meson
self-interactions and tensor terms. These additional terms not only
increase the number of two-loop integrals, but also complicate the
meson propagators, due to the short-range and many-body physics
mentioned earlier. The loop expansion is carried out to the two-loop
level, and the resulting two-loop integrals are separated into
long-range and short-range physics. The short-range contributions
are expressed in forms that already appear in the lagrangian. Since
the coefficients of these terms have yet to be determined, they are
just redefined, thus incorporating these new contributions into the
mean field energy (as in \cite{ref:Mc07}). As a result, they are
already present in the one-loop QHD calculation. The long-range
physics must be calculated explicitly. In this work, we fit the
two-loop energy to empirical properties of nuclear matter. Then we
compare parameter sets constructed at the two-loop level with those
of previous works obtained at the mean field level and consider the
naturalness of the parameters.

\section{Theory}

\subsection{Isoscalar Meson Self-Interactions and Mixing}

In this section, we investigate the effects of nonlinear meson self-interactions (including
isoscalar meson mixing) on the loop expansion.
Here we will keep terms up to order $\nu=4$ in the isoscalar meson sector of the QHD lagrangian
\cite{ref:Fu97,ref:Fu98}
(i.e.,\ $\kappa_{3}$, $\kappa_{4}$, $\eta_{1}$, $\eta_{2}$, and $\zeta_{0} \neq 0$), and let
$\alpha_{1} = \alpha_{2} = 0$.
Since we will calibrate the parameters using nuclear matter properties only
(and not finite nuclei), we want
to keep the number of parameters at a manageable level.
Therefore, the baryon and pion parts of the lagrangian are unchanged from \cite{ref:Mc07}
(except for the addition of the two-pion--nucleon vertex):
\begin{eqnarray} {\cal L} & = &
-{\psibar}\left[\gamma_{\mu}\left(\partial_{\mu} -
ig_{V}V_{\mu}\right) - i\frac{g_{A}}{f_{\pi}}\gamma_{\mu}\gamma_{5}
\partial_{\mu}\underline{\pi} 
+ \frac{1}{2f_{\pi}^{2}}\gamma_{\mu}\left[\underline{\pi},\partial_{\mu}\underline{\pi}\right]
+ \left(M-g_{S}\phi\right)\right]\psi \nonumber \\
& &
- \frac{1}{2}\left(\partial_{\mu}\pi_{a}\right)^{2} - \frac{1}{2}m_{\pi}^{2}\pi_{a}^{2}
 \ , \label{eqn:lagrangian}
\end{eqnarray}
which follows from the {\it chirally invariant\/} lagrangian in \cite{ref:Fu97} by retaining
only the relevant (lowest-order) terms in the pion fields.\footnote{In Appendix \ref{sec:app},
we discuss the role of the tensor coupling to the neutral vector meson at the two-loop level.}
The neutral scalar and vector meson lagrangian is now
\begin{eqnarray}
{\cal L}_{\phi\omega} & = & -\frac{1}{2}\left(\partial_{\mu}\phi\right)^{2} - m_{S}^{2}\phi^{2}
\left(\frac{1}{2}+\frac{\kappa_{3}}{3!}\frac{g_{S}\phi}{M} +
{ {{\kappa}_4 } \over {4!} } { {g^2_S {\phi}^2} \over {M^2} } \right)
- \frac{1}{4}V_{\mu\nu}V_{\mu\nu} \nonumber \\
& & -\frac{1}{2}m_{V}^{2}V_{\mu}V_{\mu}\left(1+\eta_{1}\frac{g_{S}\phi}{M}
+\frac{\eta_{2}}{2}\frac{g_{S}^{2}\phi^{2}}{M^{2}}\right)
+ \frac{1}{4!}\zeta_{0}g_{V}^{2}\left(V_{\mu}V_{\mu}\right)^{2} \ ,
\end{eqnarray}
\noindent where
$V_{\mu\nu}=\partial_{\mu}V_{\nu}-\partial_{\nu}V_{\mu}$ and
$\underline{\pi}=\frac{1}{2}\pi_{a}\cdot\tau_{a}$. Here $\psi$ are
the fermion fields and $\phi$, $V_{\mu}$, and $\pi_{a}$ are the
meson fields (isoscalar-scalar, isoscalar-vector, and
isovector-pseudoscalar, respectively).
The heavy meson fields are also chiral scalars.
Note that in this work, the conventions of \cite{ref:Wa04} are used.
As in \cite{ref:Mc07}, we are not making a chiral expansion in powers of the pion mass and
include it for kinematical purposes only.

The generating functional is defined in the usual way by
\begin{eqnarray}
Z[j,J_{\mu}] & \equiv & \exp\left\{iW[j,J_{\mu}]/\hbar\right\} \nonumber \\
& = & {\cal N}^{-1}\int D({\psibar})D(\psi)D(\phi)D(V_{\mu})D(\pi_{a}) \nonumber \\
& & \times \exp\left\{\frac{i}{\hbar} \int d^{4}x\left[{\cal L}(x) +
j(x)\phi(x) + J_{\mu}(x)V_{\mu}(x)\right]\right\}\ , \label{eqn:3}
\end{eqnarray}
\noindent where $\cal N$ is the normalization factor (in effect, the
vacuum subtraction, which is also an expansion in loops), $j(x)$ and $J_{\mu}(x)$ are the external
sources corresponding to the meson fields $\phi$ and $V_{\mu}$,
respectively, and the connected generating functional is $W[j,J_{\mu}]$.
The path integrals of interest here are those
over the {\it isoscalar quantum fluctuation fields} $\sigma$ and $\tilde{\eta}_{\mu}$
(the others are discussed in \cite{ref:Mc07}).
With $\eta_{1}$, $\eta_{2} \neq 0$, the integral is bilinear in form, or
\begin{eqnarray}
\lefteqn{I=\int D(\sigma)D(\tilde{\eta}_{\mu})
\exp\left\{i\int d^{4}x\left[\frac{1}{2}\,\sigma\Delta_{S}^{-1}\sigma
+ \frac{1}{2}\,\tilde{\eta}_{\mu}D_{\mu\nu}^{-1}\tilde{\eta}_{\nu} \right.\right.} & &
\nonumber \\[5pt]
& & \left.\left. {} - m_{V}^{2}\left(\frac{\eta_{1}g_{S}}{M}V_{\mu}^{0}
+\frac{\eta_{2}g_{S}^{2}}{M^{2}}\phi_{0}V_{\mu}^{0}\right)\sigma\tilde{\eta}_{\mu}
+ u\sigma +U_{\mu}\tilde{\eta}_{\mu}
\right]\vphantom{\int} \right\} \ ,
\end{eqnarray}

\noindent where the meson propagators in momentum space are
\begin{eqnarray}
\Delta_{S}^{-1}(k) & = & k^{2} + m_{S}^{2}\left(1 + \kappa_{3}\frac{g_{S}\phi_0}{M}
+ \frac{\kappa_{4}}{2}\,\frac{g^2_{S}\phi_0^2}{M^2}\right)
-m_{V}^{2}\frac{\eta_{2}}{2}\frac{g_{S}^{2}}{M^{2}}V_{\mu}^{0}V_{\mu}^{0} \ ,
\\[5pt]
{\cal D}_{\mu\nu}^{-1}(k) & = & \left[k^{2}
+ m_{V}^{2}\left(1 + \eta_{1}\frac{g_{S}\phi_0}{M} + \frac{\eta_{2}}{2}\frac{g_{S}^{2}\phi^{2}_{0}}{M^{2}}\right)
- \frac{\zeta_{0}}{6}g_{V}^{2}V_{\lambda}^{0}V_{\lambda}^{0}\right]\delta_{\mu\nu}
- k_{\mu}k_{\nu} - \frac{\zeta_{0}}{3}g_{V}^{2}V_{\mu}^{0}V_{\nu}^{0} \ . \nonumber \\
\end{eqnarray}

To simplify the equations, we introduce a five-component notation for the meson fields and
their corresponding auxiliary sources \cite{ref:Ch77}:
\begin{equation}
\underline{s} = \left(\begin{array}{c} \sigma \\ \tilde{\eta}_{\mu} \end{array}\right) \ , \quad
\underline{t} = \left(\begin{array}{c} u \\ U_{\mu} \end{array}\right) \ ,
\end{equation}

\noindent where $s_{\alpha}$ and $t_{\alpha}$ have $\alpha = 0,1, \ldots ,4$.  We can then rewrite the path integral as
\begin{eqnarray}
\lefteqn{\int D(s_{\alpha})
\exp\left\{i\int d^{4}x\left[\frac{1}{2}\,s_{\alpha}A_{\alpha\beta}^{-1}s_{\beta}
+s_{\alpha'}t_{\alpha'} \right]\right\} } & & \nonumber \\[5pt]
& = & \exp\left\{-\frac{1}{2}\tr\;
\ln\left[A^{-1}\right]\right\} \exp\left\{-\frac{i}{2}\,
t_{\alpha}A_{\alpha\beta}t_{\beta}\right\} \ ,
\end{eqnarray}

\noindent where the matrix $A^{-1}$ is
\begin{equation}
A^{-1}(k) = \left(\begin{array}{cc} \Delta_{S}^{-1}(k) & \eta V_{\nu}^{0} \\[5pt]
\eta V_{\mu}^{0} & {\cal D}_{\mu\nu}^{-1}(k) \end{array}\right) \ ,
\end{equation}

\noindent and
\begin{equation}
\eta \equiv - m_{V}^{2}\left(\frac{\eta_{1}g_{S}}{M}+\frac{\eta_{2}g_{S}^{2}}{M^{2}}\phi_{0}\right) \ .
\end{equation}
If $\eta_{1} = \eta_{2} = 0$, then $A^{-1}$ is diagonal,
there is no mixing, and the system reduces to the non-mixing case, a simple example
of which was discussed in \cite{ref:Mc07}.

Now we want to solve for the components of $A(k)$ using $A(k)A^{-1}(k) = 1$:
\begin{eqnarray}
1 & = & \left(\begin{array}{cc} A_{00} & A_{0\mu} \\[5pt]
 A_{\nu' 0} & A_{\nu'\mu} \end{array}\right)
\left(\begin{array}{cc} \Delta_{S}^{-1}(k) & \eta V_{\nu}^{0} \\[5pt]
\eta V_{\mu}^{0} & {\cal D}_{\mu\nu}^{-1}(k) \end{array}\right) \ .
\end{eqnarray}

\noindent This leads to the following four equations:
\begin{eqnarray}
A_{00}\left[\Delta_{S}^{-1}(k)\right] + A_{0\mu}\eta V_{\mu}^{0} & = & 1 \label{eqn:AB}
\ ,\\[5pt]
A_{00}\eta V_{\nu}^{0} + A_{0\mu}\left[{\cal D}_{\mu\nu}^{-1}(k)\right]
& = & 0 \label{eqn:AA} \ , \\[5pt]
A_{\nu' 0}\left[\Delta_{S}^{-1}(k)\right] + A_{\nu'\mu}\eta V_{\mu}^{0}
& = & 0 \label{eqn:AE} \ , \\[5pt]
A_{\nu' 0}\eta V_{\nu}^{0} + A_{\nu'\mu}\left[{\cal D}_{\mu\nu}^{-1}(k)\right]
& = & \delta_{\nu\nu'} \ .
\label{eqn:AF}
\end{eqnarray}

\noindent Solving these, we get ($V^{2}=V_{\mu}^{0}V_{\mu}^{0}$)
\begin{eqnarray}
A_{00} & = & \left(k^{2}+M_{V}^{2}-\frac{\zeta_{0}}{6}g_{V}^{2}V^{2}\right)
\left[\left(k^{2}+M_{V}^{2}-\frac{\zeta_{0}}{2}g_{V}^{2}V^{2}\right)\right. \nonumber \\[5pt]
& & \times \left. \left.\left(M_{V}^{2}-\frac{\zeta_{0}}{6}g_{V}^{2}V^{2}\right)
-\frac{\zeta_{0}}{3}g_{V}^{2}\left(k \cdot V\right)^{2}\right] \right/ C \ ,
\label{eqn:AK}
\end{eqnarray}

\begin{equation}
A_{0\mu} = - \eta\left.\left(k^{2}+M_{V}^{2}-\frac{\zeta_{0}}{6}g_{V}^{2}V^{2}\right)
\left[\left(M_{V}^{2}-\frac{\zeta_{0}}{6}g_{V}^{2}V^{2}\right)V_{\mu}^{0}
+\left(k\cdot V\right)k_{\mu}\right] \right/ C \ ,
\label{eqn:AL}
\end{equation}

\begin{equation}
A_{\nu'0} = - \eta\left(k^{2}+M_{V}^{2}-\frac{\zeta_{0}}{6}g_{V}^{2}V^{2}\right)
\left[\left(M_{V}^{2}-\frac{\zeta_{0}}{6}g_{V}^{2}V^{2}\right)V_{\nu'}^{0}
+\left(k\cdot V\right)k_{\nu'}\right] / C \ ,
\end{equation}

\begin{eqnarray}
A_{\nu'\mu} & = & \left(\left(k^{2}+M_{S}^{2}\right)\left\{\left[\left(k^{2}
+M_{V}^{2}-\frac{\zeta_{0}}{2}g_{V}^{2}V^{2}\right)
\left(M_{V}^{2}-\frac{\zeta_{0}}{6}g_{V}^{2}V^{2}\right)
-\frac{\zeta_{0}}{3}g_{V}^{2}\left(k \cdot V\right)^{2}\right]\delta_{\mu\nu'}\right.\right.
\nonumber \\[5pt]
& & + \left(k^{2}+M_{V}^{2}-\frac{\zeta_{0}}{2}g_{V}^{2}V^{2}\right)k_{\mu}k_{\nu'}
+\frac{\zeta_{0}}{3}g_{V}^{2}\left(M_{V}^{2}
-\frac{\zeta_{0}}{6}g_{V}^{2}V^{2}\right)V_{\mu}^{0}V_{\nu'}^{0} \nonumber \\[5pt]
& & \left.{}+\frac{\zeta_{0}}{3}g_{V}^{2}\left(k \cdot V\right)
\left(k_{\mu}V_{\nu'}^{0}+k_{\nu'}V_{\mu}^{0}\right)\right\} \nonumber \\[5pt]
& & -\eta^{2}\left\{\left[\left(M_{V}^{2}-\frac{\zeta_{0}}{6}g_{V}^{2}V^{2}\right)V^{2}
+\left(k \cdot V\right)^{2}\right]\delta_{\mu\nu'}+V^{2}k_{\mu}k_{\nu'} \right. \nonumber \\[5pt]
& & \left.\left.{} -\left(M_{V}^{2}-\frac{\zeta_{0}}{6}g_{V}^{2}V^{2}\right)V_{\mu}^{0}V_{\nu'}^{0}
-\left(k \cdot V\right)\left(k_{\mu}V_{\nu'}^{0}+k_{\nu'}V_{\mu}^{0}\right)\right\}\right) / C \ ,
\label{eqn:AH}
\end{eqnarray}

\noindent where
\begin{eqnarray}
C & \equiv & \left(k^{2}+M_{V}^{2}-\frac{\zeta_{0}}{6}g_{V}^{2}V^{2}\right) \nonumber \\[5pt]
& & \times \left\{\left(k^{2}+M_{S}^{2}\right)
\left[\left(k^{2}+M_{V}^{2}-\frac{\zeta_{0}}{2}g_{V}^{2}V^{2}\right)
\left(M_{V}^{2}-\frac{\zeta_{0}}{6}g_{V}^{2}V^{2}\right)
-\frac{\zeta_{0}}{3}g_{V}^{2}\left(k \cdot V\right)^{2}\right]\right. \nonumber \\[5pt]
& & \left. {} - \eta^{2}\left[\left(M_{V}^{2}-\frac{\zeta_{0}}{6}g_{V}^{2}V^{2}\right)V^{2} 
+ \left(k \cdot V\right)^{2}\right]\right\} \ ,
\label{eqn:AD}
\end{eqnarray}

\noindent and
\begin{eqnarray}
M_{S}^{2} & \equiv & m_{S}^{2}\left(1 + \kappa_{3}\frac{g_{S}\phi_{0}}{M}
+ \frac{\kappa_{4}}{2}\,\frac{g^2_{S}\phi^2_{0}}{M^2}\right)
-m_{V}^{2}\frac{\eta_{2}}{2}\frac{g_{S}^{2}V^{2}}{M^{2}} \ ,
\label{eqn:AI} \\[5pt]
M_{V}^{2} & \equiv & m_{V}^{2}\left(1 + \eta_{1}\frac{g_{S}\phi_{0}}{M}
+ \frac{\eta_{2}}{2}\frac{g_{S}^{2}\phi_{0}^{2}}{M^{2}}\right) \ .
\label{eqn:AJ}
\end{eqnarray}

The corresponding path integral in the vacuum subtraction has a similar structure,
except with vanishing
classical meson fields; therefore, $A$ is replaced by $A_{0}$, or
\begin{equation}
A_{0}(k) = \left(\begin{array}{cc} \Delta_{S}^{0}(k)  & 0 \\
                                   0 & {\cal D}_{\mu\nu}^{0}(k) \end{array}\right) \ ,
\end{equation}

\noindent where $\Delta_{S}^{0}$ and ${\cal D}_{\mu\nu}^{0}$ are the free propagators for the
scalar and neutral vector mesons, respectively:
\begin{eqnarray}
\Delta_{S}^{0}(k) & = &
\frac{1}{k^{2}+m_{S}^{2}-i\epsilon} \ , \label{eqn:prop1} \\  {\cal
D}_{\mu\nu}^{0}(k) & = &
\frac{1}{k^{2}+m_{V}^{2}-i\epsilon} \left(\delta_{\mu\nu}
+ \frac{k_{\mu}k_{\nu}}{m_{V}^{2}}\right) \ . \label{eqn:prop2}
\end{eqnarray}

The pion self-couplings do not contribute to the energy density if there is no pion condensate.
Therefore, we simply ignore the pure pion terms in the following calculation.
The remaining generating functional is
\begin{eqnarray}
\lefteqn{Z\left[j,J_{\mu}\right] = {\cal N}^{-1}
\exp\left\{\frac{i}{\hbar}\int d^{4}x\left[{\cal L}_{0}(x) + j(x)\phi_{0}(x)
+ J_{\mu}(x)V_{\mu}^{0}(x)\right]\right\}} & & \nonumber \\[5pt]
& & {} \times \exp\left\{\int d^{4}x\int\frac{d^{4}k}{(2\pi)^{4}}
\left(\tr \; \ln \left[G_{F}^{0}(k)G_{H}^{-1}(k)\right]
- \frac{1}{2}\,\tr \; \ln \left[A_{0}(k)A^{-1}(k)\right]\right)\right\} \nonumber \\[5pt]
& & \times \left[\!\!\left[ \exp\left\{i\hbar^{1/2}\int d^{4}x
\left[\frac{i\delta}{\delta \xi(x)}\right]
\left(g_{S}\left[\frac{-i\delta}{\delta u(x)}\right]
+ig_{V}\gamma_{\mu}\left[\frac{-i\delta}{\delta U_{\mu}(x)}\right]
\right.\right.\right.\right. \nonumber \\[5pt]
& & \left. \quad {}+i\frac{g_{A}}{2f_{\pi}}\gamma_{\mu}\gamma_{5}\partial_{\mu}\left[
\frac{-i\delta}{\delta \zeta_{a}(x)}\right]
\cdot\tau_{a}\right)\left[\frac{-i\delta}{\delta \bar{\xi}(x)}\right] \nonumber \\[5pt]
& & {} -i\hbar^{1/2}\int d^{4}x\left(m_{S}^{2}
\left(\frac{\kappa_{3}g_{S}}{6M}+\frac{\kappa_{4}g_{S}^{2}}{6M^{2}}\phi_{0}\right)
\left[\frac{-i\delta}{\delta u(x)}\right]^{3} \right. \nonumber \\
& & \left. {} +
m_{V}^{2}\left(\frac{\eta_{1}g_{S}}{2M}+\frac{\eta_{2}g_{S}^{2}}{2M^{2}}\phi_{0}\right)
\left[\frac{-i\delta}{\delta
u(x)}\right]\left[\frac{-i\delta}{\delta U_{\mu}(x)}\right]\left[
\frac{-i\delta}{\delta U_{\mu}(x)}\right]\right. \nonumber \\[5pt]
& & \left. \quad {}-\frac{\zeta_{0}}{6}g_{V}^{2}V_{\mu}^{0}\left[
\frac{-i\delta}{\delta U_{\mu}(x)}\right]
\left[\frac{-i\delta}{\delta U_{\nu}(x)}\right]\left[
\frac{-i\delta}{\delta U_{\nu}(x)}\right]
+m_{V}^{2}\frac{\eta_{2}g_{S}^{2}}{2M^{2}}V_{\mu}^{0}\left[\frac{-i\delta}{\delta u(x)}\right]^{2}
\left[\frac{-i\delta}{\delta U_{\mu}(x)}\right]\right) \nonumber \\[5pt]
& & {} +\hbar\int d^{4}x\left( \frac{1}{24}\, \zeta_{0}g_{V}^{2}
\left[\frac{-i\delta}{\delta U_{\mu}(x)}\right]\left[\frac{-i\delta}{\delta U_{\mu}(x)}\right]
\left[\frac{-i\delta}{\delta U_{\nu}(x)}\right]\left[\frac{-i\delta}{\delta U_{\nu}(x)}\right]
-m_{S}^{2}\frac{\kappa_{4}g_{S}^{2}}{24M^{2}}\left[\frac{-i\delta}{\delta u(x)}\right]^{4} \right.
\nonumber \\[5pt]
& & \left.\left. {} - m_{V}^{2}\frac{\eta_{2}g_{S}^{2}}{4M^{2}}\left[\frac{-i\delta}{\delta u(x)}\right]^{2}
\left[\frac{-i\delta}{\delta U_{\mu}(x)}\right]\left[\frac{-i\delta}{\delta U_{\mu}(x)}\right]
\right)\right\} \nonumber \\
& & {}\times \exp\left\{i\int d^{4}x\int d^{4}y \left[\frac{1}{2}\,
t_{\alpha}A_{\alpha\beta}(x-y)t_{\beta} \right. \right. \nonumber \\[5pt]
& & \left.\left.\left.\left. \quad {}+ \frac{1}{2}\,
\zeta_{a}(x)\Delta_{\pi}^{ab}(x-y)\zeta_{b}(y)
- \bar{\xi}(x)G_{H}(x-y)\xi(y)\right]\right\}
\right]\!\!\right]_{\mathrm{sources}\ =\ 0} \ .
\end{eqnarray}

\noindent Here $\zeta_{\alpha}$ is the pion source and $\xi$ is the nucleon source.
We define the contribution at the mean field level as
\begin{eqnarray}
{\cal L}_{0} & = & - \frac{1}{2}\,
m_{V}^{2}V_{\mu}^{0}V_{\mu}^{0}\left(1 +
\eta_{1}\frac{g_{S}\phi_{0}}{M} +
\frac{\eta_{2}}{2}\frac{g_{S}^{2}\phi_{0}^{2}}{M^{2}}\right) -
m_{S}^{2}\phi_{0}^{2}\left(\frac{1}{2} +
\frac{\kappa_{3}}{3!}\frac{g_{S}\phi_{0}}{M} +
\frac{\kappa_{4}}{4!}\frac{g^2_{S}\phi^2_{0}}{M^2} \right) \nonumber
\\[5pt]
& & {} + \frac{1}{4!}\,
\zeta_{0}g_{V}^{2}\left(V_{\mu}^{0}V_{\mu}^{0}\right)^{2} \ ,
\end{eqnarray}
and the pion and fermion propagators are given by
\begin{eqnarray}
\Delta_{\pi}^{ab}(k) & = &
\frac{1}{k^{2}+m_{\pi}^{2}-i\epsilon}\,\delta_{ab}\ , \label{eqn:prop3}
\label{eqn:PION} \\[5pt]
G_{F}^{0}(k) & = & \frac{-1}{i{\not\! k}+M-i\epsilon} \ , \\[3pt]
G_{H}(k) & = & \frac{-1}{i{\not\!
k}-ig_{V}\gamma_{\mu}V_{\mu}^{0}+
\left(M-g_{S}\phi_{0}\right)}\ .
\end{eqnarray}

It is straightforward to acquire the connected generating functional at the one-loop order
\begin{equation}
W_{1} = -i\hbar\int d^{4}x \int\frac{d^{4}k}{(2\pi)^{4}}
\left\{\tr \; \ln \left[G_{F}^{0}(k)G_{H}^{-1}(k)\right]
- \frac{1}{2}\, \tr \; \ln \left[A_{0}(k)A^{-1}(k)\right]\right\}\ ,
\end{equation}

\noindent and therefore
\begin{eqnarray}
\Gamma^{(1)}\left[\phi_{e},V_{e}^{0}\right] & = & \int d^{4}x\left\{{\cal L}_{0}
+ i\hbar\int\frac{d^{4}k}{(2\pi)^{4}}
\left(\vphantom{\int} \tr \; \ln \left[G_{F}^{0}(k)G_{H}^{-1}(k)\right] \right.\right. \nonumber
\\[5pt]
& & \left. \left. {}- \frac{1}{2}\, \tr \; \ln \left[A_{0}(k)A^{-1}(k)\right]\right)\right\} \ .
\end{eqnarray}

\noindent The baryon term is identical to the simple case in
\cite{ref:Mc07}. To treat the isoscalar terms, we first consider the
identity
\begin{equation}
\tr \; \ln \left[A_{0}(k)A^{-1}(k)\right] \equiv \ln \det\left[A_{0}(k)A^{-1}(k)\right] \ ,
\end{equation}

\noindent since it is easy to acquire the determinants:
\begin{eqnarray}
\det\left[A_{0}(k)\right]
& = & \left[m_{V}^{2}\left(k^{2} + m_{S}^{2}\right)\left(k^{2}
+ m_{V}^{2}\right)^{3}\right]^{-1} \ , \\[5pt]
\det\left[A^{-1}(k)\right]
& = & \left(k^2+M_{V}^{2}-\frac{\zeta_{0}}{6}g_{V}^{2}V_{e}^{2}\right)^{2}
\left\{\left(k^{2}+M_{S}^{2}\right)\left[\left(k^{2}
+M_{V}^{2}-\frac{\zeta_{0}}{6}g_{V}^{2}V_{e}^{2}\right)\right.\right. \nonumber \\[5pt]
& &  \left. \times \left(M_{V}^{2}-\frac{\zeta_{0}}{2}g_{V}^{2}V_{e}^{2}\right)
+\frac{\zeta_{0}}{3}g_{V}^{2}\left[k^{2}V_{e}^{2}-\left(k \cdot V_{e}\right)^{2}\right]\right] \nonumber \\[5pt]
& & \left. \quad {} - \eta^{2}\left[\left(M_{V}^{2}-\frac{\zeta_{0}}{6}g_{V}^{2}V_{e}^{2}\right)V_{e}^{2}
+ \left(k \cdot V_{e}\right)^{2}\right]\right\} \ ,
\end{eqnarray}

\noindent where $M_{S}^{2}$ and $M_{V}^{2}$ are defined by Eqs.\ (\ref{eqn:AI})
and (\ref{eqn:AJ}), and
${V_{\mu}^{0}}^{(1)} = V^{e}_{\mu} = i\delta_{\mu 4}V_{e}$.

Apparently, $\det\left[A^{-1}(k)\right]$ is an analytic function of $\phi_{e}$ and $V_{e}$, and
\begin{equation}
\det\left[A^{-1}(k)\right]\left(\phi_{e}=0,V_{e}=0\right) = \det\left[A_{0}^{-1}(k)\right] \ .
\end{equation}

\noindent Therefore, the meson contribution to the effective action can be
categorized by the general form
\begin{equation}
\mathop{{\sum}'}_{i,\,j\, =\, 0}^{\infty}
\alpha_{ij}\phi^{i}\left(V_{\mu}V_{\mu}\right)^{j}\ ,
\end{equation}

\noindent where the prime indicates that $i+j > 0$.
However, terms of this type already appear in the
general QHD lagrangian; since the coefficients of these terms have yet to be determined,
these meson contributions can be absorbed
into the preexisting terms. As a result, they don't need to be separately considered.

Now we want to determine the connected generating functional at the two-loop level, $W_{2}$.
For this, we have to make a Legendre transformation to get the effective action.
The classical meson fields are approximated as \cite{ref:Mc07}
\begin{equation}
\left(\begin{array}{c} \phi_{0}(x) \\[5pt] V_{\mu}^{0}(x) \end{array}\right)
= \left(\begin{array}{c} \phi_{e}(x) \\[5pt] V^{e}_{\mu}(x) \end{array}\right)
+ \int d^{4}y A(x-y)\left(\begin{array}{c}
\left.\displaystyle{\frac{\delta W_{1}}{\delta\phi_{0}(y)}}\right|_{e}
\\[12pt] \left.\displaystyle{\frac{\delta W_{1}}{\delta V_{\mu}^{0}(y)}}\right|_{e}
\end{array}\right) + O(\hbar^{2}) \ .
\end{equation}

\noindent One can verify that the tadpole diagrams in $W_{2}$ eventually cancel,
and so we are left with only the one-particle irreducible diagrams:
\begin{eqnarray}
\lefteqn{\Gamma^{(2)}[\phi_{e},V_{\mu}^{e}]
= \int d^{4}x\,{\cal L}_{0}\left[\phi_{e}(x),V_{\mu}^{e}(x)\right]} & &  \nonumber \\[5pt]
& & {}- i\hbar \int d^{4}x\int\frac{d^{4}k}{(2\pi)^{4}}
\left\{\tr \; \ln \left[G_{F}^{0}(k)G_{H}^{-1}(k)\right]
- \frac{1}{2}\, \tr \; \ln \left[A_{0}(k)A^{-1}(k)\right]\right\} \nonumber \\[5pt]
& & {}+\hbar^{2}\int d^{4}x \int\frac{d^{4}k}{(2\pi)^{4}}\int\frac{d^{4}q}{(2\pi)^{4}}
\left\{\frac{g_{S}^{2}}{2}A_{00}(k-q) \tr\left[G_{H}(k)G_{H}(q)\right] \right. \nonumber \\[5pt]
& & \quad {} -\frac{g_{V}^{2}}{2}A_{\mu\nu}(k-q)
\tr\left[G_{H}(k)\gamma_{\mu}G_{H}(q)\gamma_{\nu}\right]
\nonumber \\[5pt]
& & \quad {}-\frac{g_{A}^{2}}{2f_{\pi}^{2}}\, \Delta^{ab}_{\pi}(k-q)
\tr\left[G_{H}(k)(\not\! k - \not\! q)\gamma_{5}\frac{\tau_a}{2}
G_{H}(q)(\not\! k - \not\! q)\gamma_{5}\frac{\tau_b}{2}\right]
\nonumber \\[5pt]
& & \quad {}+ ig_{S}g_{V}A_{0\mu}(k-q) \tr\left[\gamma_{\mu}G_{H}(k)G_{H}(q)\right]
\nonumber \\[5pt]
& & \quad {}+3\left(\kappa_{3}\frac{g_{S}m_{S}^{2}}{6M}
+ \kappa_{4}\frac{g^2_{S}m_{S}^{2}\phi_0}{6M^2}\right)^{2}
A_{00}(k)A_{00}(q)A_{00}(k-q)
\nonumber \\[5pt]
& & \quad {}-\kappa_{4}\frac{g^{2}_{S}m_{S}^{2}}{8M^2}A_{00}(k)A_{00}(q)
\nonumber \\[5pt]
& & \quad {} +\frac{1}{4}\, \eta^{2}\left[A_{00}(k-q)A_{\mu\nu}(k)A_{\mu\nu}(q)
+ 2A_{0\mu}(k)A_{0\nu}(q)A_{\mu\nu}(k-q)\right] \nonumber \\[5pt]
& & \quad {}+\frac{1}{2}\, \eta\left(\kappa_{3}\frac{g_{S}m_{S}^{2}}{M}
+ \kappa_{4}\frac{g^2_{S}m_{S}^{2}\phi_0}{M^2}\right)A_{00}(k-q)A_{0\mu}(k)A_{0\mu}(q)
\nonumber \\[5pt]
& & \quad {}-\eta_{2}\frac{g_{S}^{2}m_{V}^{2}}{4M^{2}}
\left[A_{00}(k)A_{\mu\mu}(q)+2A_{0\mu}(k)A_{0\mu}(q)\right] \nonumber \\[5pt]
& & \quad {}+\frac{1}{24}\, \zeta_{0}g_{V}^{2}\left[A_{\nu\nu}(k)A_{\mu\mu}(q)
+2A_{\mu\nu}(k)A_{\mu\nu}(q)\right] \nonumber \\[5pt]
& & \quad {}+\frac{\eta_{2}^{2}g_{S}^{4}}{2M^{4}}\, m_{V}^{4}
\left[A_{00}(k)A_{00}(q)V_{\mu}^{e}A_{\mu\nu}(k-q)V_{\nu}^{e}
+2A_{00}(k)V_{\mu}^{e}A_{\mu 0}(q)A_{0\nu}(k-q)V_{\nu}^{e}\right] \nonumber \\[5pt]
& & \quad {}-m_{V}^{2}\frac{\eta_{2}\zeta_{0}g_{S}^{2}g_{V}^{2}}{6M^{2}}
\left[A_{0\lambda}(k)A_{\lambda\mu}(q)V_{\mu}^{e}V_{\nu}^{e}A_{\nu 0}(k-q)
+2A_{0\lambda}(k)A_{\lambda 0}(q)V_{\mu}^{e}A_{\mu\nu}(k-q)V_{\nu}^{e}\right] \nonumber \\[5pt]
& & \quad {}+\frac{1}{36}\, \zeta_{0}^{2}g_{V}^{4}\left[A_{\mu\nu}(k)A_{\mu\nu}(q)
V_{\lambda}^{e}A_{\lambda\lambda'}(k-q)V_{\lambda'}^{e} \right. \nonumber \\[5pt]
& & \left.\left. \qquad\qquad\qquad {}+2V_{\mu}^{e}A_{\mu\nu}(k)A_{\nu\lambda}(k-q)
A_{\lambda\lambda'}(q)V_{\lambda'}^{e}\right] - \mathrm{VEV}
\vphantom{\int} \right\} \ ,
\end{eqnarray}

\noindent where VEV is the vacuum subtraction.
Notice that there are now four distinct fermion loops at the two-loop level. These reduce to the
familiar three loops from the non-mixing case when $\eta_{1} = \eta_{2} = 0$.

With dimensional regularization, we can shift momentum variables to remove the dependence on the
vector field from the fermion propagators and thereby replace $G_{H}(k)$ by $G^{*}(k)$, where
\begin{eqnarray}
G^{*}(k) & = & \frac{-1}{i{\not\! k}+M-g_{S}\phi_{0}} \nonumber \\[5pt]
& = & \left(i{\not\! k}-M^{*}\right)\left[\frac{1}{k^{2}+{M^{*}}^{2}-i\epsilon}
- \frac{i\pi}{E^{*}(k)}\delta[k_{4} - E^{*}(k)]\theta(k_{F}-|\vec{k}|)\right] \nonumber \\[5pt]
& \equiv & G_{F}^{*}(k) + G_{D}^{*}(k) \ .
\label{eqn:BB3}
\end{eqnarray}
Here $M^{*}=M-g_{S}\phi_{0}$.
We can then separate each of the fermion loops [using $G^{*}(k) = G_{F}^{*}(k) + G_{D}^{*}(k)$]
into three contributions:
an exchange term, a Lamb-shift term, and a vacuum-fluctuation term.
The exchange terms will be evaluated explicitly later.
For now, let us consider the Lamb-shift term for the fermion loop involving scalar--vector mixing:
\begin{eqnarray}
\lefteqn{\int\frac{d^{4}k}{(2\pi)^{4}} \int\frac{d^{4}q}{(2\pi)^{4}}\,
A_{0\mu}(k-q) \tr\left[\gamma_{\mu}G_{F}^{*}(k)G_{D}^{*}(q)\right]} & & \nonumber \\[5pt]
& \propto & \int\frac{d^{4}k}{(2\pi)^{4}} \int\frac{d^{4}q}{(2\pi)^{4}}\,
\delta\left(q_{4} - E^{*}(q)\right)
\theta\left(k_{F} - |\mathbf{q}|\right)\frac{i\pi}{E^{*}(q)}
\left[\frac{M^{*}(k+q)_{\mu}}{k^{2}+{M^{*}}^{2}}\right] \nonumber \\[5pt]
& & {} \times \left[\left(k-q\right)^{2}+{M_{V}^{*}}^{2}\right]
\left(\left[\left(k-q\right)\cdot V\right](k-q)_{\mu} + {M_{V}^{*}}^{2}V_{\mu}\right)
\nonumber \\[5pt]
& & \times \left\{ \vphantom{\int}\!\left[\left(k-q\right)^{2}+M_{S}^{2}\right]
\left[\left(k-q\right)^{2}+{M_{V}^{*}}^{2}\right]\right. \nonumber \\[5pt]
& & \times \left(\left[\left(k-q\right)^{2}+{M_{V}^{*}}^{2}
+\frac{\zeta_{0}}{3}g_{V}^{2}V^{2}\right]{M_{V}^{*}}^{2}
-\frac{\zeta_{0}}{3}g_{V}^{2}\left[\left(k-q\right)\cdot V\right]^{2}\right) \nonumber \\[5pt]
& & \left. {}- \eta^{2}\left(\left[\left(k-q\right)^{2}+{M_{V}^{*}}^{2}\right]{M_{V}^{*}}^{2}V^{2}
+\left[\left(k-q\right)^{2}+{M_{V}^{*}}^{2}+\frac{\zeta_{0}}{3}g_{V}^{2}V^{2}\right]
\left[\left(k-q\right)\cdot V\right]^{2}\right)\right\}^{-1} \nonumber \\[5pt]
& \propto & \int\frac{d^{4}q}{(2\pi)^{4}} \int\frac{d^{4}p}{(2\pi)^{4}}
\frac{\theta\left(k_{F} - |\mathbf{q}|\right)}{E^{*}(q)}
\frac{\delta\left(q_{4} - E^{*}(q)\right)}
{\left[p^{2}+2\left(p\cdot q\right)+q^{2}+{M^{*}}^{2}\right]} \nonumber \\[5pt]
& & {} \times \left[p^{2}+{M_{V}^{*}}^{2}\right]
\left[\left(p\cdot V\right)\left(p^{2}+2p\cdot q+{M_{V}^{*}}^{2}\right)
+ 2{M_{V}^{*}}^{2}\left(q\cdot V\right)\right] \nonumber \\[5pt]
& & \times \left\{\left(p^{2}+M_{S}^{2}\right)\left(p^{2}+{M_{V}^{*}}^{2}\right)
\left[\left(p^{2}+{M_{V}^{*}}^{2}+\frac{\zeta_{0}}{3}g_{V}^{2}V^{2}\right){M_{V}^{*}}^{2}
-\frac{\zeta_{0}}{3}g_{V}^{2}\left(p\cdot V\right)^{2}\right]\right. \nonumber \\[5pt]
& & \left. {} - \eta^{2}\left[\left(p^{2}+{M_{V}^{*}}^{2}\right){M_{V}^{*}}^{2}V^{2}
+\left(p^{2}+{M_{V}^{*}}^{2}+\frac{\zeta_{0}}{3}g_{V}^{2}V^{2}\right)
\left(p\cdot V\right)^{2}\right]\right\}^{-1} \nonumber \\[5pt]
& \propto & \int\frac{d^{4}q}{(2\pi)^{4}}
\int\frac{d^{4}p}{(2\pi)^{4}}\frac{\theta\left(k_{F} - |\mathbf{q}|\right)}{E^{*}(q)}
\,\delta\left(q_{4} - E^{*}(q)\right)\sum_{i\, =\, 0}^{\infty}\sum_{j\, =\, 0}^{\infty}
F_{ij}\left(p^{2},q^{2},V^{2},M^{*}\right) \nonumber \\[5pt]
& & \times \left(p\cdot q\right)^{i}\left(p\cdot V\right)^{j}
\left\{\left(p\cdot V\right)\left[p^{2}+2\left(p\cdot q\right)
+{M_{V}^{*}}^{2}\right]+2{M_{V}^{*}}^{2}
\left(q\cdot V\right)\right\} \nonumber \\[5pt]
& \propto & \int\frac{d^{4}q}{(2\pi)^{4}} \frac{\theta\left(k_{F} - |\mathbf{q}|\right)}{E^{*}(q)}
\,\delta\left(q_{4} - E^{*}(q)\right) \widetilde{F}\left(M^{*},V^{2},q\cdot V\right) \ ,
\end{eqnarray}

\noindent where
\begin{equation}
{M_{V}^{*}}^{2} \equiv M_{V}^{2} - \frac{\zeta_{0}}{6}g_{V}^{2}V^{2}\ ,
\end{equation}

\noindent $k_{F}$ is the Fermi momentum, and
$E^{*}(q) \equiv (\mathbf{q}^{2} + {M^{*}}^{2})^{1/2}$.
We have made the substitution $p_{\mu} = (k - q)_{\mu}$,
which is allowed if we keep $q$ fixed in the $k$ integral.
The function $F$ comes from the fractional integrand and is an analytic function of the
variables $p^{2}$, $q^{2}$, $V^{2}$, and $M^{*}$.
The function $\widetilde{F}$ comes from the integral over $p$ and is analytic in
the variables $M^{*}$, $V^{2}$, and $q\cdot V$, since $p_{\mu}$ is always spacelike.
The on-shell condition $q_{4} = E^{*}(q)$ is used to eliminate the
dependence on $q^{2}$ in the last line.

Remember that $V_{\mu}^{0} = iV_{0}\delta_{\mu 4}$.
Thus, after integrating over $q_{4}$, the remaining expression is an analytic function
of $E^{*}(q)$.
The final result will be complicated, but should be explicitly dependent on some types of
baryon densities.
These terms can be directly related to terms in the effective lagrangian that are some
combination of the nucleon fields (with some derivatives) multiplied
by some Lorentz-invariant polynomials in the mean meson fields;
as a result, they can be parametrized by local terms in the effective lagrangian.
A simple case was illustrated in \cite{ref:Mc07}.

The term with the pion propagator is the same as in the non-mixing case.
The fermion loops with $A_{00}$
and $A_{\mu\nu}$ will, in the case of $\eta_{1} = \eta_{2} = 0$, reduce to the equivalent loops with the pure
neutral scalar and vector mesons from the non-mixing case.
We expect that they will behave similarly in both the mixing and non-mixing cases, with the
former simply leading to more general local combinations of the fields.
{\it Thus the Lamb-shift terms can be parametrized by terms already present in the lagrangian
(before truncation).}

The vacuum-fluctuation terms and all the pure meson loops have no explicit density dependence,
so all the conclusions of the previous analysis \cite{ref:Mc07} are still applicable here.
The short-range physics which is contained in the Lamb-shift and vacuum-fluctuation terms
is absorbed in preexisting terms in the QHD lagrangian.
The long-range physics arising from the exchange term must be explicitly calculated;
these integrals are considered in the following section.

\subsection{Two-loop Integral Formulas}

\noindent We first consider the integral with the propagator from Eq.\ (\ref{eqn:AK}).
Therefore, one can write
\begin{eqnarray}
{\cal E}^{(2)}_{\phi - EX}
& = & -\frac{g_{S}^{2}}{2}\int\frac{d^{4}k}{(2\pi)^{4}} \int\frac{d^{4}q}{(2\pi)^{4}}\,
A_{00}(k-q) \tr\left[G_{D}^{*}(k)G_{D}^{*}(q)\right] \nonumber \\[5pt]
& = & \frac{\gamma g_{S}^{2}}{32\pi^{4}}\int_{0}^{k_{F}}
\frac{|\mathbf{k}|^{2}d|\mathbf{k}|}{E^{*}(k)}
\int_{0}^{k_{F}}\frac{|\mathbf{q}|^{2}d|\mathbf{q}|}{E^{*}(q)} \int_{-1}^{1}d(\cos\theta)
\nonumber \\[5pt]
& & \times \left[E^{*}(k)E^{*}(q)-|\mathbf{k}||\mathbf{q}|\cos\theta+{M^{*}}^{2}\right]
\nonumber \\[5pt]
& & \times \left[2E^{*}(k)E^{*}(q)-2|\mathbf{k}||\mathbf{q}|\cos\theta
-2{M^{*}}^{2}+{M_{V}^{*}}^{2}\right] \nonumber \\[5pt]
& & \times \left\{\left[2E^{*}(k)E^{*}(q)-2|\mathbf{k}||\mathbf{q}|\cos\theta
-2{M^{*}}^{2}+{M_{V}^{*}}^{2}
{} +\frac{\zeta_{0}}{3}g_{V}^{2}V_{0}^{2}\right]{M_{V}^{*}}^{2} \right. \nonumber \\[5pt]
& & \left.\left. -\frac{\zeta_{0}}{3}g_{V}^{2}V_{0}^{2}\left[E^{*}(k)-E^{*}(q)\right]^{2} \right\}
\right/ B \ ,
\end{eqnarray}

\noindent where
\begin{eqnarray}
B & \equiv & \left[2E^{*}(k)E^{*}(q)-2|\mathbf{k}||\mathbf{q}|
\cos\theta-2{M^{*}}^{2}+M_{S}^{2}\right]
\nonumber \\[5pt]
& & \times \left[2E^{*}(k)E^{*}(q)-2|\mathbf{k}||\mathbf{q}|
\cos\theta-2{M^{*}}^{2}+{M_{V}^{*}}^{2}\right] \nonumber \\[5pt]
& & \times \left(\left[2E^{*}(k)E^{*}(q)-2|\mathbf{k}||\mathbf{q}|
\cos\theta-2{M^{*}}^{2}+{M_{V}^{*}}^{2}
+\frac{\zeta_{0}}{3}g_{V}^{2}V_{0}^{2}\right]{M_{V}^{*}}^{2} \right. \nonumber \\[5pt]
& & \left.\qquad {} -\frac{\zeta_{0}}{3}g_{V}^{2}V_{0}^{2}
\left[E^{*}(k)-E^{*}(q)\right]^{2} \right) \nonumber \\[5pt]
& & {} + \eta^{2}V_{0}^{2}\left(\left[2E^{*}(k)E^{*}(q)-2|\mathbf{k}||\mathbf{q}|
\cos\theta-2{M^{*}}^{2}+{M_{V}^{*}}^{2}\right]
{M_{V}^{*}}^{2}\right. \nonumber \\[5pt]
& & {} -\left[2E^{*}(k)E^{*}(q)-2|\mathbf{k}||\mathbf{q}|\cos\theta-2{M^{*}}^{2}+{M_{V}^{*}}^{2}
+\frac{\zeta_{0}}{3}g_{V}^{2}V_{0}^{2}\right] \nonumber \\[5pt]
& & \quad \left. {} \times \left[E^{*}(k)-E^{*}(q)\right]^{2}\right) \ .
\label{eqn:den}
\end{eqnarray}

Now we consider the integral with the propagator from Eq.\ (\ref{eqn:AH}); as a result, we get
\begin{eqnarray}
{\cal E}^{(2)}_{V - EX} & = &
\frac{g_{V}^{2}}{2}\int\frac{d^{4}k}{(2\pi)^{4}} \int\frac{d^{4}q}{(2\pi)^{4}}\,
A_{\mu\nu}(k-q) \tr\left[\gamma_{\mu}G_{D}^{*}(k)\gamma_{\nu}G_{D}^{*}(q)\right] \nonumber \\[5pt]
& = & \frac{\gamma g_{V}^{2}}{32\pi^{4}}\int_{0}^{k_{F}}
\frac{|\mathbf{k}|^{2}d|\mathbf{k}|}{E^{*}(k)}
\int_{0}^{k_{F}}\frac{|\mathbf{q}|^{2}d|\mathbf{q}|}{E^{*}(q)} \int_{-1}^{1}d(\cos\theta)
\nonumber \\[5pt]
& & \times \left[\left[2E^{*}(k)E^{*}(q)-2|\mathbf{k}||\mathbf{q}|
\cos\theta-2{M^{*}}^{2}+M_{S}^{2}\right] \right.
\nonumber \\[5pt]
& & \times \left\{2\left(\left[2E^{*}(k)E^{*}(q)-2|\mathbf{k}||\mathbf{q}|
\cos\theta-2{M^{*}}^{2}+{M_{V}^{*}}^{2}
+\frac{\zeta_{0}}{3}g_{V}^{2}V_{0}^{2}\right]{M_{V}^{*}}^{2}\right.\right. \nonumber \\[5pt]
& & \left. -\frac{\zeta_{0}}{3}g_{V}^{2}V_{0}^{2}
\left[E^{*}(k)-E^{*}(q)\right]^{2}\right) \nonumber \\[5pt]
& & \times \left[E^{*}(k)E^{*}(q)-|\mathbf{k}||\mathbf{q}|\cos\theta-2{M^{*}}^{2}\right]
\nonumber \\[5pt]
& & \left. {} +\frac{\zeta_{0}}{3}g_{V}^{2}V_{0}^{2}{M_{V}^{*}}^{2}
\left[E^{*}(k)E^{*}(q)+|\mathbf{k}||\mathbf{q}|\cos\theta+{M^{*}}^{2}\right]\right\} \nonumber \\ 
& & -\eta^{2}V_{0}^{2}\left(2\left\{\left[E^{*}(k)-E^{*}(q)\right]^{2}-{M_{V}^{*}}^{2}\right\}
\left[E^{*}(k)E^{*}(q)-|\mathbf{k}||\mathbf{q}|\cos\theta-2{M^{*}}^{2}\right] \right. \nonumber \\[5pt]
& & \left.\left.\left. -{M_{V}^{*}}^{2}\left[E^{*}(k)E^{*}(q)+|\mathbf{k}||\mathbf{q}|\cos\theta+{M^{*}}^{2}\right]\right)\right]
\right/ B \ . 
\end{eqnarray}

Lastly, we consider the integral with the mixed propagator from Eq.\ (\ref{eqn:AL}),
which becomes
\begin{eqnarray}
{\cal E}^{(2)}_{M - EX} & = & -ig_{S}g_{V}\int\frac{d^{4}k}{(2\pi)^{4}} \int\frac{d^{4}q}{(2\pi)^{4}}\,
A_{0\mu}(k-q) \tr\left[\gamma_{\mu}G_{D}^{*}(k)G_{D}^{*}(q)\right] \nonumber \\[5pt]
& = & -\frac{\gamma g_{S}g_{V}\eta M^{*}V_{0}}{16\pi^{4}}{M_{V}^{*}}^{2}
\int_{0}^{k_{F}}\frac{|\mathbf{k}|^{2}d|\mathbf{k}|}{E^{*}(k)}
\int_{0}^{k_{F}}\frac{|\mathbf{q}|^{2}d|\mathbf{q}|}{E^{*}(q)} \int_{-1}^{1}d(\cos\theta)
\nonumber \\[7pt]
& & \times \left[E^{*}(k)+E^{*}(q)\right]
\left.\left[2E^{*}(k)E^{*}(q)-2|\mathbf{k}||\mathbf{q}|
\cos\theta-2{M^{*}}^{2}+{M_{V}^{*}}^{2}\right] \right/ B \ .
\end{eqnarray}

\noindent The formula for the pion contribution to the two-loop energy
density is the same as in \cite{ref:Mc07}.

\section{Discussion}

In the previous section, we showed that the short- and long-range dynamics of the two-loop
contributions can be separated even in the presence of various nonlinearities in the isoscalar
meson fields.
The short-range physics can be expressed as a series of terms that are already present
in the underlying lagrangian; thus, they are implicitly contained in the undetermined
coefficients.
The long-range exchange contributions must be calculated explicitly.
The question remains what effect these new contributions have on the naturalness assumption.

Naturalness \cite{ref:Ge93,ref:Ma84,ref:Fr96,ref:Fr97,ref:Fu97,ref:Fu98} states that once all the dimensional
and combinatorial factors in a given term of the lagrangian are removed, what is left is a
dimensionless constant of order unity.
To show that naturalness is not violated by the inclusion of the two-loop exchange integrals,
we will show that sets of natural parameters exist that adequately reproduce the empirical
properties of symmetric nuclear matter at equilibrium \cite{ref:Fu99}.

\begin{table}
\begin{center}
\begin{tabular}{|c|c|c|c|c|c|c|} \hline
             & Q2 \cite{ref:Fu97} & M1A          & C1 \cite{ref:Fu97} & M2A          & M3A              & M4A             \\ \hline
$m_{S}/M$    & $\zz 0.54268$   & $\zz 0.5400\xx$ & $\zz 0.53874$   & $\zz 0.5400\xx$ & $\zz 0.5400\xx$  & $\zz 0.5400\xx$ \\ \hline
$m_{V}/M$    & $\zz 0.83280$   & $\zz 0.8328\xx$ & $\zz 0.83280$   & $\zz 0.8328\xx$ & $\zz 0.8328\xx$  & $\zz 0.8328\xx$ \\ \hline
$g_{S}/4\pi$ & $\zz 0.78661$   & $\zz 0.80762$   & $\zz 0.77756$   & $\zz 0.91847$   & $\zz 0.84973$    & $\zz 0.83407$   \\ \hline
$g_{V}/4\pi$ & $\zz 0.97202$   & $\zz 0.93485$   & $\zz 0.98486$   & $\zz 1.15837$   & $\zz 0.98992$    & $\zz 0.96368$   \\ \hline
$\eta_{1}$   & ---             & ---             & $\zz 0.29577$   & $\zz 1.9207\xx$ & $\zz 0.45972$    & $\zz 0.25290$   \\ \hline
$\eta_{2}$   & ---             & ---             & ---             & ---             & ---              & $\zz 0.19238$   \\ \hline
$\kappa_{3}$ & $\zz 1.7424\xx$ & $\zz 1.2072\xx$ & $\zz 1.6698\xx$ & $\zz 4.3207\xx$ & $\zz 2.6308\xx$  & $\zz 2.1978\xx$ \\ \hline
$\kappa_{4}$ & $   -8.4836\xx$ & $   -1.8824\xx$ & ---             & ---             & $    -5.6463\xx$ & $   -5.3909\xx$ \\ \hline
$\zeta_{0}$  & $   -1.7750\xx$ & $\zz 9.9968\xx$ & ---             & ---             & $\zz 8.9212\xx$  & $\zz 9.0313\xx$ \\ \hline
\end{tabular}
\caption{Parameter sets used in this work compared with one-loop parameter sets at the
same level of truncation.}
\label{tab:1}
\end{center}
\end{table}

We conduct parameter fits at
different levels of truncation in the underlying lagrangian, as in \cite{ref:Fu97}, but here we
include the two-loop exchange contributions.
Note that the two-loop portion of the energy density contains no new parameters.
In this work, we fit the parameters to bulk properties of nuclear matter:
${\cal E}/\rho_{B}-M=-16.10$ MeV, $k_{F}=1.30$ fm$^{-1}$, $M^{*}/M=0.610$, and
compressibility $K=250\pm 50$ MeV. A least-squares fit was conducted to these four
pieces of data with respective weights of $0.0015$, $0.002$, $0.0015$, and $0.08$.
The fitting was performed using the downhill simplex method,
with nuclear equilibrium maintained throughout the procedure.
Table \ref{tab:1} lists the results of this fitting for a number of different levels of truncation
in the lagrangian (M1A, M2A, M3A, and M4A)---defined below---along with some corresponding parameter sets
determined at the one-loop level with the same truncation (Q2 and C1 \cite{ref:Fu97})
for comparison.
We hold the meson masses fixed in the new sets ($m_{S}/M=0.5400$ and $m_{V}/M=0.8328$),
since there are only four empirical input data; they accurately reproduce
the reduced meson mass in \cite{ref:Fu02}. We let $M=939$ MeV.

\begin{itemize}
\item The mean field lagrangian for the M1A (and Q2) set includes cubic
and quartic scalar field self-couplings, as well as the quartic
vector field self-coupling. In this case there is no mixing in the
meson propagators ($\eta=0$); as a result, there are only three two-loop
exchange integrals ($A_{0\mu}$ drops out). 

\item The M2A (and C1) set
includes a cubic scalar field self-coupling in addition to the
third-order mixing term. This case contains mixing in the meson
propagators through the inclusion of the $\eta_{1}$ term; therefore,
there are four exchange integrals to calculate. 

\item Two other mixing
sets were constructed (M3A and M4A); however, no comparable mean
field set exists in the literature. These sets have the same terms
as M1A plus $\eta_{1}\neq 0$ and $\eta_{1}$, $\eta_{2}\neq 0$,
respectively. Both sets give rise to four two-loop integrals as in
the M2A case.
\end{itemize}

\begin{table}
\begin{center}
\begin{tabular}{|c|c|c|c|c|c|c|} \hline
                      & Q2 \cite{ref:Fu97} & M1A      & C1 \cite{ref:Fu97} & M2A     & M3A     & M4A     \\ \hline
${\cal E}/\rho_{B}-M$ &$-16.13$            &$-16.10$  &$ -16.19 $          &$-16.11$ &$-16.10$ &$-16.10$ \\ \hline
$k_{F}$               & 1.303              & 1.300    & 1.293              & 1.306   & 1.300   & 1.300   \\ \hline
$M^{*}/M$             & 0.614              & 0.610    & 0.532              & 0.611   & 0.610   & 0.610   \\ \hline
$g_{V}V_{0}$          & 292                & 234      & 255                & 239     & 231     & 230     \\ \hline
$K$                   & 279                & 270      & 304                & 288     & 235     & 247     \\ \hline
\end{tabular}
\caption{Nuclear matter properties for the parameter sets in Table \ref{tab:1}. The values of
${\cal E}/\rho_{B}-M$, $g_{V}V_{0}$, and $K$ are in MeV. The Fermi wavenumber is in fm$^{-1}$
and $M^{*}/M$ is dimensionless.}
\label{tab:2}
\end{center}
\end{table}
\begin{table}
\begin{center}
\begin{tabular}{|c|c|c|c|c|c|c|} \hline
                                                       & Q2 \cite{ref:Fu97} & M1A          & C1 \cite{ref:Fu97}
& M2A         & M3A           & M4A           \\ \hline
${\cal E}_{\phi-EX}^{(2)}\vphantom{\displaystyle\sum}$ & $\zz 38.77$        & $ \zz 33.51$ & $\zz 26.73$
& $\zz 21.26$ & $\zz 31.39$   & $\zz 32.99$   \\ \hline
${\cal E}_{V-EX}^{(2)}\vphantom{\displaystyle\sum}$    & $-28.32$           & $-15.49$     & $-23.47$
& $-18.50$    & $-15.85$      & $-15.67$      \\ \hline
${\cal E}_{M-EX}^{(2)}\vphantom{\displaystyle\sum}$    & ---                & ---          & $\zz\xx 3.54$
& $\zz 11.99$ & $\zz\xx 4.24$ & $\zz\xx 3.30$ \\ \hline
${\cal E}_{\pi-EX}^{(2)}\vphantom{\displaystyle\sum}$  & $\zz 12.87$        & $\zz 12.76$  & $\zz 12.73$
& $\zz 12.79$ & $\zz 12.71$   & $\zz 12.68$   \\ \hline
\end{tabular}
\caption{Size of two-loop integrals for the parameter sets in Table \ref{tab:1}.
Values are in MeV.}
\label{tab:3}
\end{center}
\end{table}

In the set M1A, the value of $\zeta_{0}$ was allowed to vary; however, $\zeta_{0}$ places a
strong constraint on the compressibility $K$.
One must also ensure that the vector field equation still has an appropriate solution at
high density. Confining $\zeta_{0}$ to positive values is sufficient to reproduce
the correct high-density behavior.
One can see from Table \ref{tab:1} that parameter sets exist at the two-loop level
in which all the parameters are natural.
{\it In all cases the properties of nuclear matter were reproduced
accurately while maintaining naturalness.}
The bulk nuclear properties for the various sets are shown in Table \ref{tab:2};
note that all the observables
in this table except $g_{V}V_{0}$ were used in the fit.

\begin{figure}
\begin{center}
\includegraphics[width=4 in]{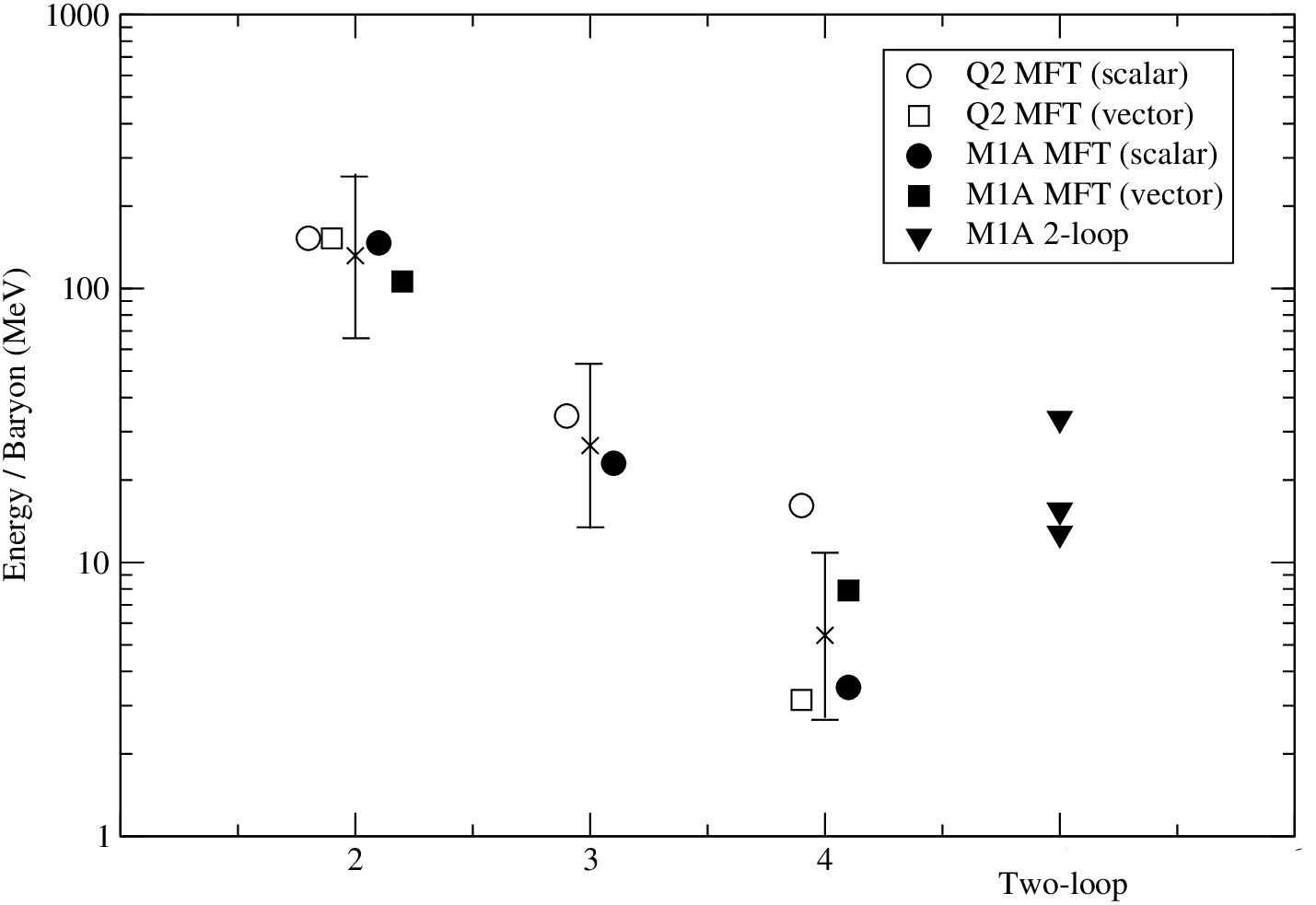}
\caption{Comparison of the magnitudes of the mean field terms in the
meson sector with the two-loop exchange integrals.  The inverted
triangles represent, from top to bottom, the scalar, vector, and
pion two-loop integrals. The abscissa denotes the order $\nu$ in the
power counting at the mean field level \cite{ref:Fu97}. \vspace{.1
in}} \label{fig:loop1}
\end{center}
\end{figure}


\begin{figure}
\begin{center}
\includegraphics[width=4 in]{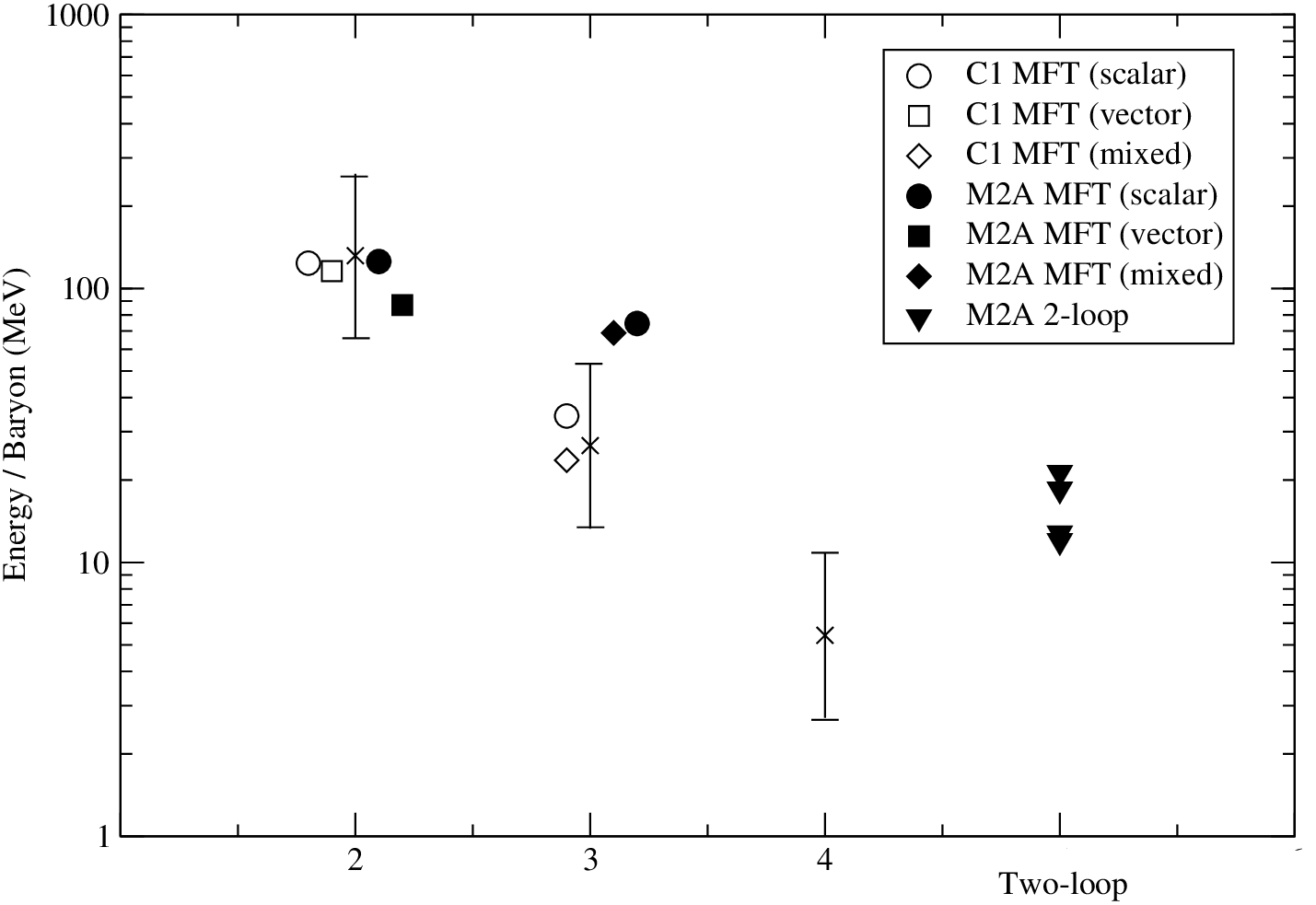}
\caption{Comparison of the magnitudes of the mean field terms in the
meson sector with the two-loop exchange integrals. The inverted
triangles represent, from top to bottom, the scalar, vector, pion,
and mixed two-loop integrals. The abscissa denotes the order $\nu$
in the power counting at the mean field level \cite{ref:Fu97}.}
\label{fig:loop2}
\end{center}
\end{figure}

\begin{figure}
\begin{center}
\includegraphics[width=4 in]{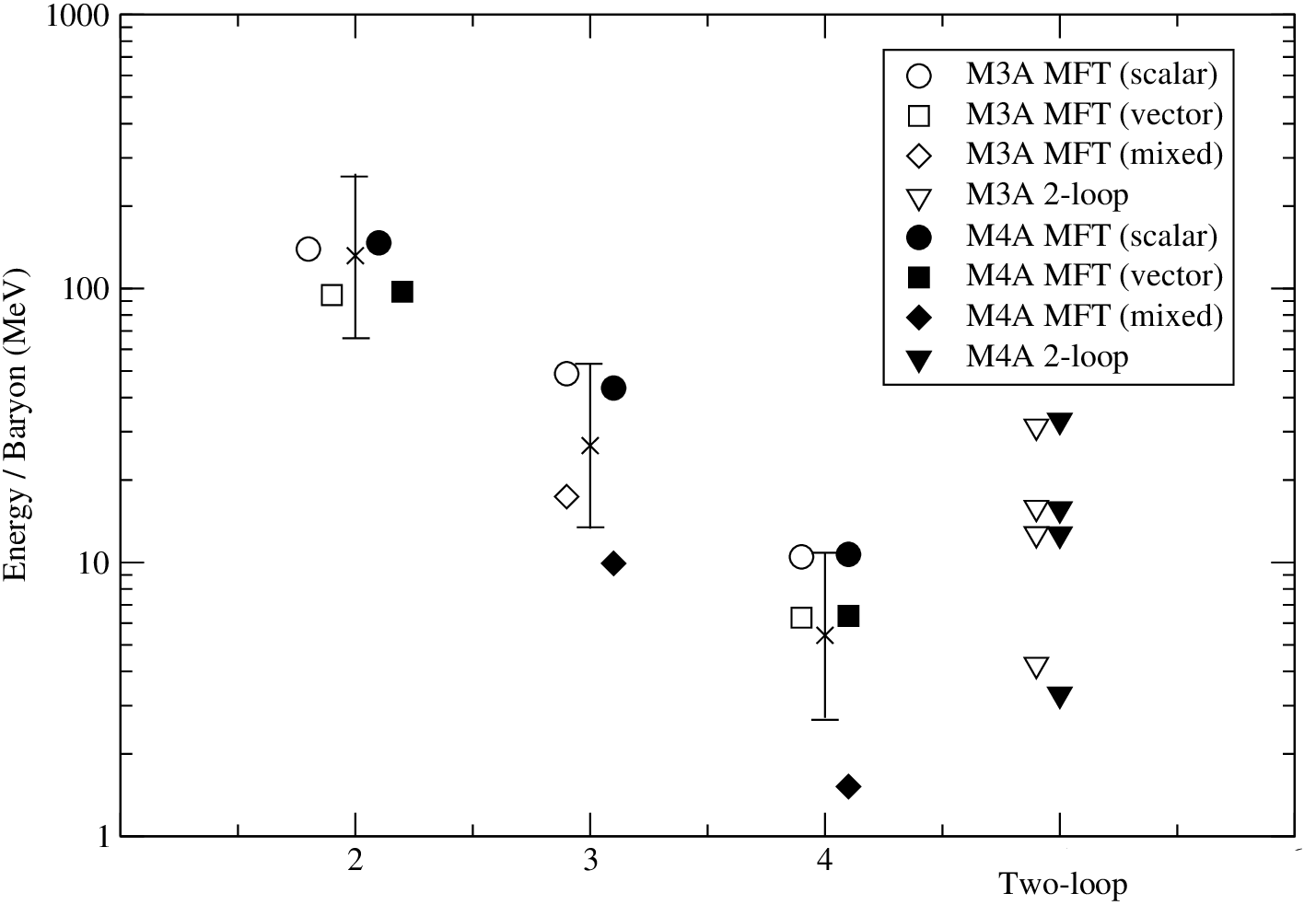}
\caption{Comparison of the magnitudes of the mean field terms in the
meson sector with the two-loop exchange integrals.  The inverted
triangles represent, from top to bottom, the scalar, vector, pion,
and mixed two-loop integrals. The abscissa denotes the order $\nu$
in the power counting at the mean field level \cite{ref:Fu97}.
\vspace{.1 in}} \label{fig:loop3}
\end{center}
\end{figure}

The size of the various two-loop integrals is shown in Table
\ref{tab:3}. We also show the size of these integrals when the
constants are taken from the Q2 and C1 sets. Note that there is some
cancellation between the scalar and vector contributions.

\begin{figure}
\begin{center}
\includegraphics[width=4 in]{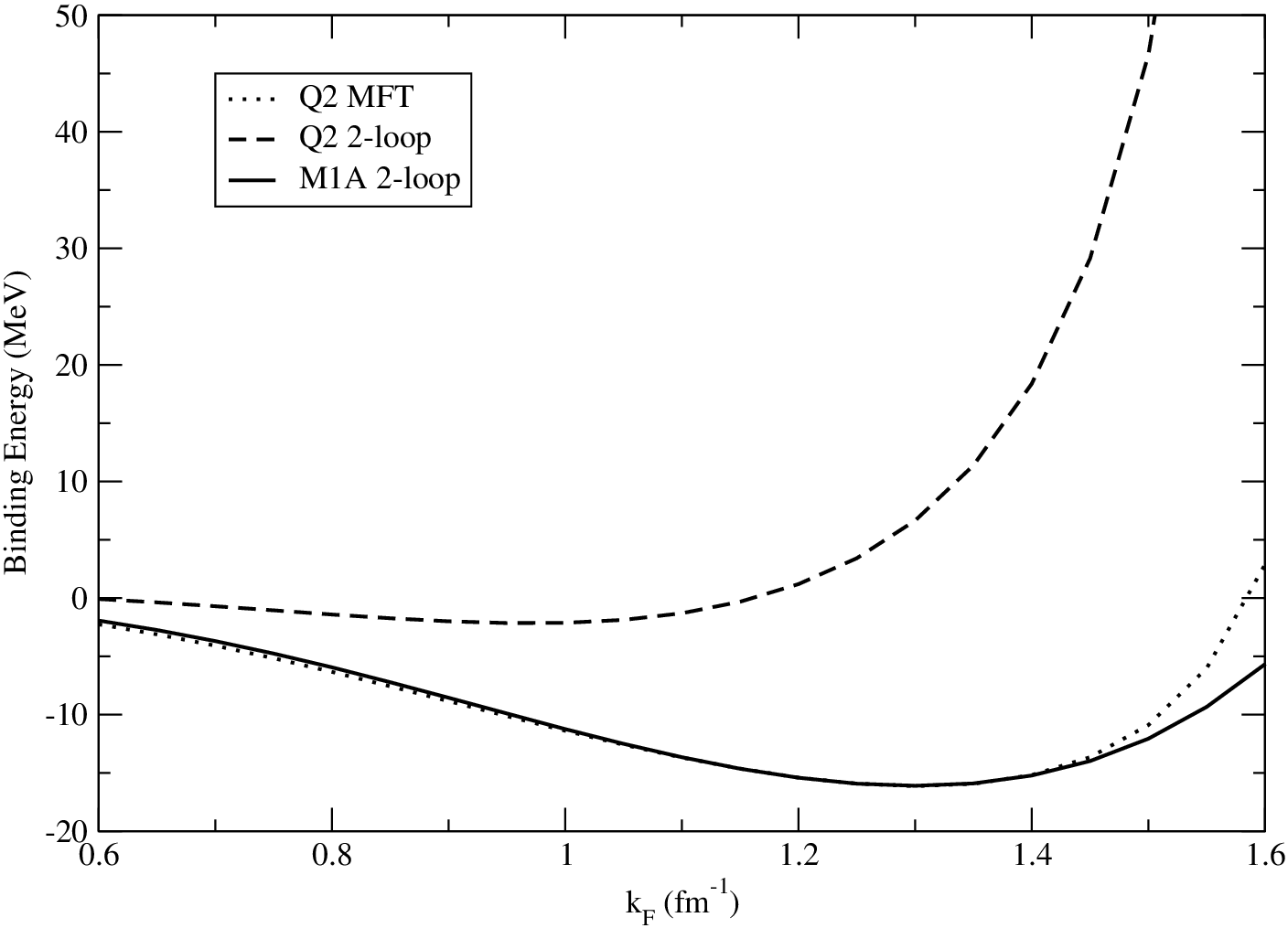}
\caption{Comparison of the nuclear binding curves for the set Q2
(both at the one-loop and two-loop level) with the set M1A
(two-loop level).\vspace{.1 in}} \label{fig:sat_cur1}
\end{center}
\end{figure}

Next, we compare the magnitude of the two-loop integrals to the
mean-field terms. In Fig.\ \ref{fig:loop1}, the mean field terms for
the Q2 and M1A sets are plotted order by order in the power
counting, with the two-loop integrals calculated using M1A. The
crosses in Fig.\ \ref{fig:loop1} represent the expected magnitude
per order of terms in the meson sector using the rules of naive
dimensional analysis (NDA) \cite{ref:Fu97}, with the chiral symmetry
breaking scale of $\Lambda = 650$ MeV.  Fig.\ \ref{fig:sat_cur1}
compares the saturation curve of M1A at the two-loop level with the
saturation curve of Q2 at the one-loop level and the two-loop
binding curve of Q2. Apparently, the net effect of the exchange terms is repulsive.

\begin{figure}
\begin{center}
\includegraphics[width=4 in]{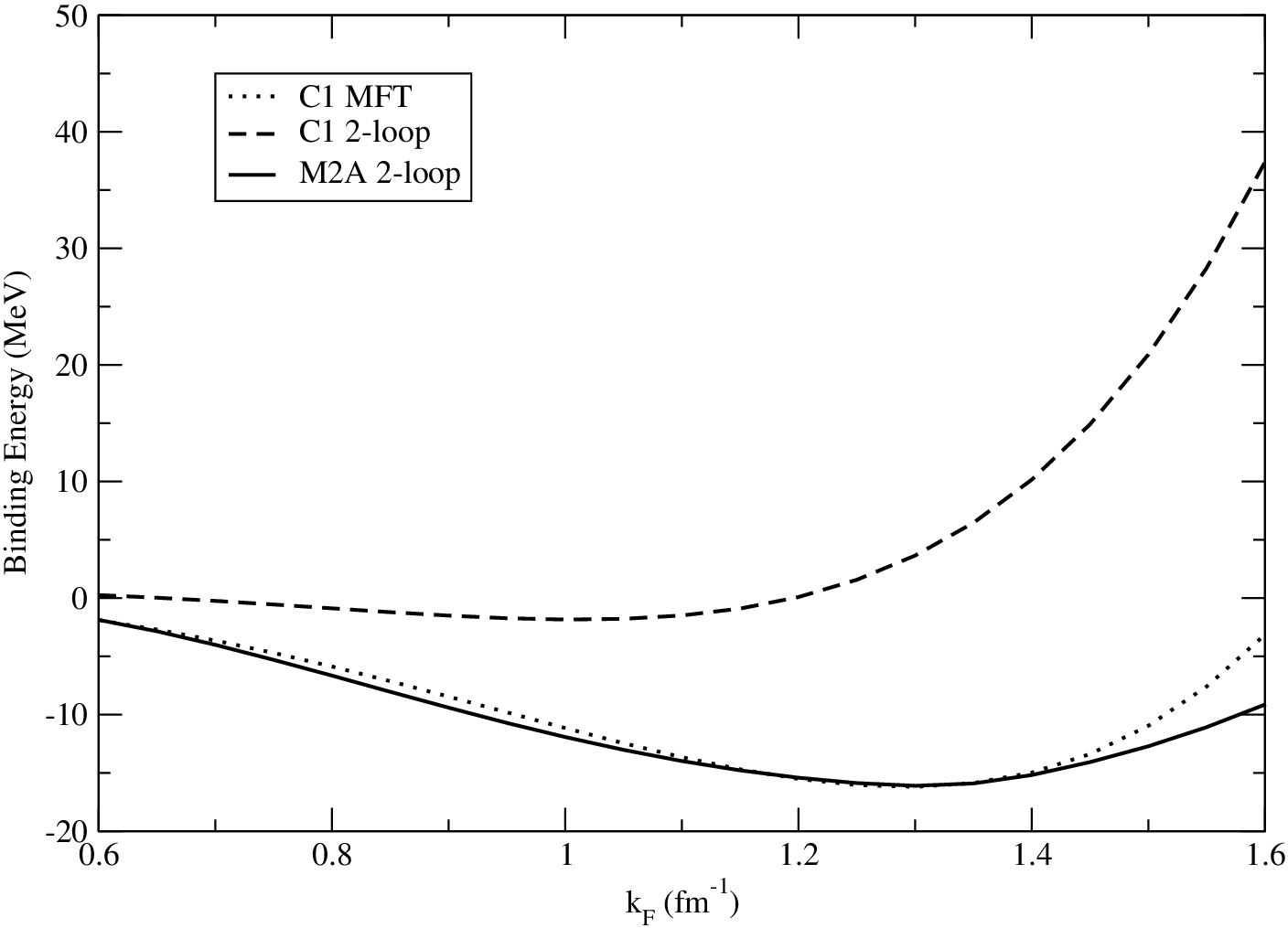}
\caption{Comparison of the nuclear binding curves for the set C1
(both at the one-loop and two-loop level) with the set M2A
(two-loop level).\vspace{.1 in}} \label{fig:sat_cur2}
\end{center}
\end{figure}

In Fig.\ \ref{fig:loop2}, the mean field terms
for the C1 and M2A sets are plotted order by order in the power
counting, with the two-loop integrals calculated using M2A. A
comparison of the saturation curve of M2A at the two-loop level with
the saturation curve of C1 at the one-loop level and the two-loop
binding curve of C1 is shown in Fig.\ \ref{fig:sat_cur2}. 

\begin{figure}
\begin{center}
\includegraphics[width=4 in]{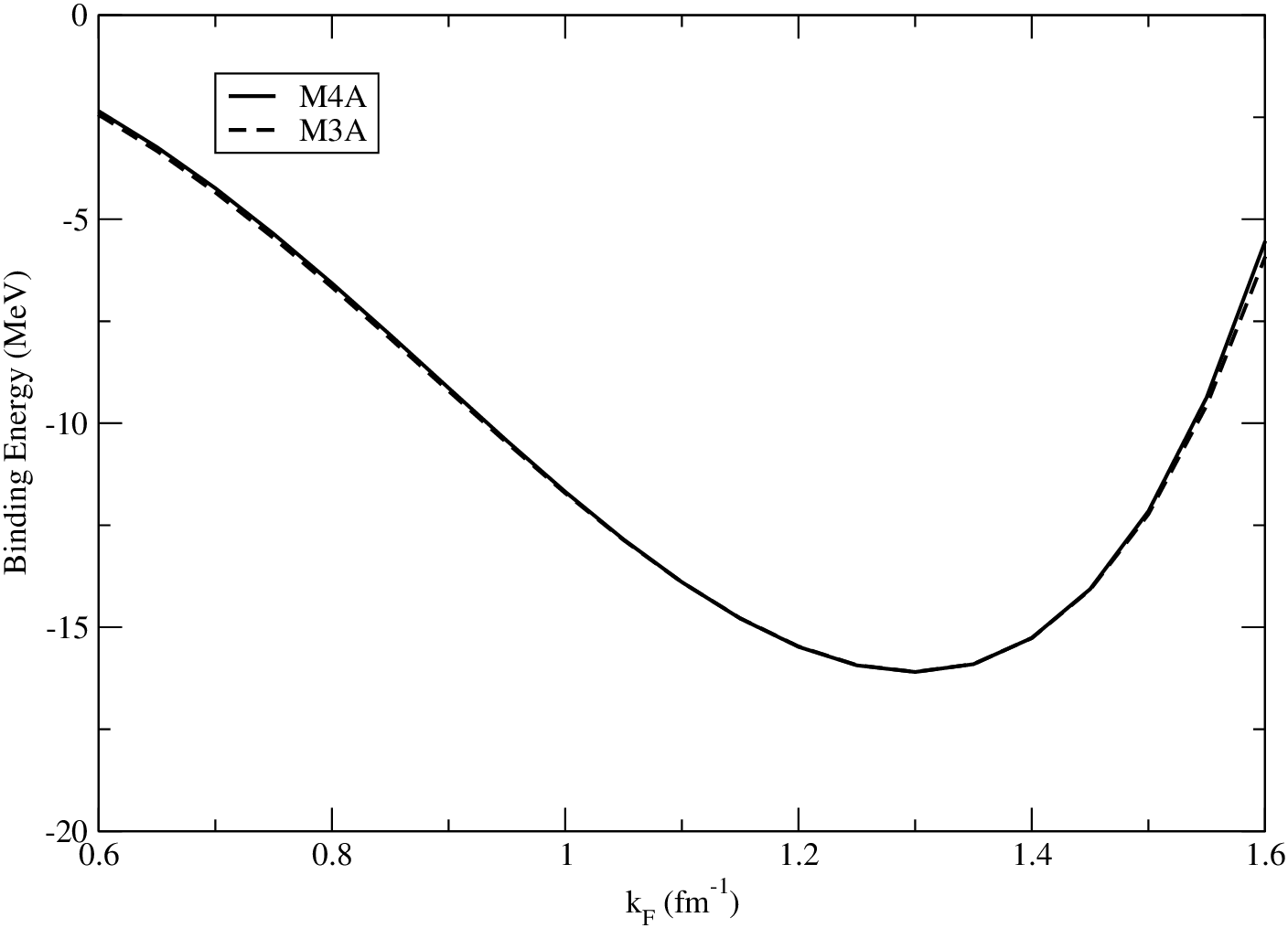}
\caption{Comparison of the saturation curves for the sets M3A
and M4A (both at the two-loop level).\vspace{.1 in}} \label{fig:sat_cur3}
\end{center}
\end{figure}

Lastly, the terms (mean field and two-loop) for both M3A and M4A are plotted
in Fig.\ \ref{fig:loop3}. Fig.\ \ref{fig:sat_cur3} is a graph of the
saturation curves for both M3A and M4A at the two-loop level.

One can see from Figs.\ \ref{fig:loop1}, \ref{fig:loop2}, and
\ref{fig:loop3} that the magnitudes of the mean field terms are
consistent with the NDA estimates to within a factor of two, which
supports the naturalness of the coupling parameters.\footnote{The determination 
of naturalness is a tricky business, and the relevant observation here is that
the hierarchy of successive terms in Figs.\ \ref{fig:loop1}, \ref{fig:loop2}, 
and \ref{fig:loop3} is maintained with the new parameter sets.
A more precise discussion of the naturalness would require an examination of the linearly
independent combinations of parameters, as in \cite{ref:Fu99}.}
Moreover, the magnitude of the two-loop terms is shown to be roughly equal to the
third order $(\nu = 3)$ in the power counting in the mean field
energy. While this is not large, they cannot be neglected in a
description of nuclear matter properties, particularly in view of
the nearly complete cancellation of scalar and vector terms at order
$\nu = 2$. Nevertheless, when the parameters are properly
renormalized to reproduce empirical equilibrium properties, exchange
corrections modify the energy only slightly, and the two-loop
binding curves for nuclear matter are similar to the one-loop
results.


Exchange interactions are two-body, so they go like $O(\rho^{2})$ in
$\cal E$. This is the same dependence as the two-body mean field
terms at $O(\nu=2)$. Because of the lack of a spin--isospin sum in
the exchange terms, the exchange terms are numerically smaller and
contribute at the same level as the $O(\nu=3)$ terms. A deeper
understanding of how loops fit into the finite-density power
counting scheme will require a look at higher-order terms in the
loop expansion, which is currently under investigation \cite{ref:Mc08}.

Exchange contributions from the the vector meson two-loop diagrams
with tensor couplings at one or both vertices were also considered.
The analytic expressions for these integrals are given in the
Appendix, and the numerical results are shown in Table \ref{tab:4}.
The vector--tensor cross terms (XT) are roughly equal to fourth
order $(\nu = 4)$ in the power counting, and the tensor--tensor
terms (TT) are negligible. This is not surprising considering that
the tensor coupling on a vertex introduces an additional factor of
$\nabla /M \approx 1/4$ (see \cite{ref:Fu97} for a description of
the power counting). A value of $f_{V}=0.5$ was used here, which is
comparable to values obtained from fits to properties of finite
nuclei \cite{ref:Fu97}. The tensor terms are small enough that their
effect on the naturalness of the parameters is negligible.

\begin{table}
\begin{center}
\begin{tabular}{|c|c|c|c|c|c|c|} \hline
                                                      & Q2 \cite{ref:Fu97} & M1A          & C1 \cite{ref:Fu97} & M2A              \\ \hline
${\cal E}_{TT-EX}^{(2)}\vphantom{\displaystyle\sum}$  & $-0.0241$           & $-0.0189$     & $-0.0210$           & $-0.0195$     \\ \hline
${\cal E}_{XT-EX}^{(2)}\vphantom{\displaystyle\sum}$  & $-3.94\xx\xx$       & $-2.95\xx\xx$ & $-3.43\xx\xx$       & $-3.12\xx\xx$ \\ \hline
\end{tabular}
\caption{Size of two-loop integrals involving the tensor coupling to the vector meson
for the parameter sets in Table \ref{tab:1}. Values are in MeV.}
\label{tab:4}
\end{center}
\end{table}

In {\it conclusion}, we have shown that the results of
\cite{ref:Mc07} are still valid when various nonlinearities in the
isoscalar meson fields are included in the lagrangian. The short-
and long-range physics at the two-loop level in the loop expansion
can be separated. The short-range (local) dynamics is parametrized
by the undetermined coefficients of the lagrangian, like those in the
nonlinear meson self-interactions, and is either removed by field redefinitions 
or fitted to empirical data. The long-range
(nonlocal) physics must be calculated explicitly. The assumption of
naturalness is shown to hold in the presence of the two-loop
exchange integrals. The size of the two-loop contributions is
roughly equal to third order in the mean field power counting. The
inclusion of tensor contributions was also considered and found to
be small.

\section*{Acknowledgements}

We thank our colleague J. D. Walecka for valuable
comments on the manuscript. This work was supported in part by the Department
of Energy under Contract No.\ DE--FG02--87ER40365.

\appendix
\section{Tensor Portion of Two-Loop Calculation}
\label{sec:app}

In this section, we present the corrections to the theory arising
from the tensor coupling to the neutral vector meson at the two-loop
level, where
\begin{equation}
{\cal L}_{T} = \frac{f_{V}g_{V}}{4M} \,
\bar{\psi}\sigma_{\mu\nu}V_{\mu\nu}\psi \ .
\end{equation}

\noindent After working out the variational derivatives, we acquire
the energy density. Now we can replace the Hartree propagator by
$G_{H} \rightarrow G^{*} = G^{*}_{F} + G^{*}_{D}$, as before. The
Lamb-shift and vacuum fluctuation terms are absorbed by the same
process as before. (The only difference here is more gamma matrices
and they don't influence the result.) Now we are left with only the
exchange terms
\begin{eqnarray}
{\cal E}^{(2)}_{TT-EX} & = & \frac{f_{V}^{2}g_{V}^{2}}{8M^{2}}\int\frac{d^{4}k}{(2\pi)^{4}}
\int\frac{d^{4}q}{(2\pi)^{4}}A_{\nu\sigma}(k-q) \nonumber \\
& & {} \times
\tr\left[\sigma_{\mu\nu}\left(k-q\right)_{\mu}G^{*}_{D}(k)
\sigma_{\rho\sigma}\left(k-q\right)_{\rho}G^{*}_{D}(q)\right] \nonumber \\
& = & -\frac{f_{V}^{2}g_{V}^{2}}{64\pi^{4}M^{2}}
\int_{0}^{k_{F}}\frac{|\mathbf{k}|^{2}d|\mathbf{k}|}{E^{*}(k)}
\int_{0}^{k_{F}}\frac{|\mathbf{q}|^{2}d|\mathbf{q}|}{E^{*}(q)} \int_{-1}^{1}d(\cos\theta)
\nonumber \\[7pt]
& & \times \left(\left[2E^{*}(k)E^{*}(q)-2|\mathbf{k}||\mathbf{q}|
\cos\theta-2{M^{*}}^{2}+M_{S}^{2}\right]\right. \nonumber \\
& & \times \left\{2\left(\left[2E^{*}(k)E^{*}(q)-2|\mathbf{k}||\mathbf{q}|
\cos\theta-2{M^{*}}^{2}+M_{V}^{2}+\frac{\zeta_{0}}{2}g_{V}^{2}V_{0}^{2}\right]\right.\right. \nonumber \\
& & \times \left. \left[M_{V}^{2}+\frac{\zeta_{0}}{6}g_{V}^{2}V_{0}^{2}\right]
-\frac{\zeta_{0}}{3}g_{V}^{2}V_{0}^{2}\left[E^{*}(k)-E^{*}(q)\right]^{2}\right) \nonumber \\
& & \times \left[E^{*}(k)E^{*}(q)-|\mathbf{k}||\mathbf{q}|
\cos\theta-{M^{*}}^{2}\right]^{2} \nonumber \\
& & +\frac{\zeta_{0}}{3}g_{V}^{2}V_{0}^{2}\left[M_{V}^{2}+\frac{\zeta_{0}}{6}g_{V}^{2}V_{0}^{2}\right] \nonumber \\
& & \times \left(\left[E^{*}(k)E^{*}(q)-|\mathbf{k}||\mathbf{q}|\cos\theta-{M^{*}}^{2}\right]
\left[2E^{*}(k)E^{*}(q)-2{M^{*}}^{2}\right] \right. \nonumber \\
& & \left.\left. -{M^{*}}^{2}\left[E^{*}(k)-E^{*}(q)\right]^{2}\right)\right\} \nonumber \\
& & -\eta^{2}V_{0}^{2}\left[2\left\{\left[E^{*}(k)-E^{*}(q)\right]^{2}
-\left(M_{V}^{2}+\frac{\zeta_{0}}{6}g_{V}^{2}V_{0}^{2}\right)\right\}\right. \nonumber \\
& & \times \left[E^{*}(k)E^{*}(q)-|\mathbf{k}||\mathbf{q}|\cos\theta-{M^{*}}^{2}\right]^{2} \nonumber \\
& & +\left(M_{V}^{2}+\frac{\zeta_{0}}{6}g_{V}^{2}V_{0}^{2}\right)
\left\{2\left[E^{*}(k)E^{*}(q)-|\mathbf{k}||\mathbf{q}|\cos\theta-{M^{*}}^{2}\right]\right. \nonumber \\
& & \left.\left.\left. \times \left[E^{*}(k)E^{*}(q)-{M^{*}}^{2}\right]-{M^{*}}^{2}\left[E^{*}(k)-E^{*}(q)\right]^{2}\right\}\right]\right)
/ B \ , \nonumber \\
{\cal E}^{(2)}_{XT-EX} & = & \frac{f_{V}g_{V}^{2}}{2M}\int\frac{d^{4}k}{(2\pi)^{4}}
\int\frac{d^{4}q}{(2\pi)^{4}}
A_{\mu\nu}(k-q) \nonumber \\
& & {} \times
\tr\left[\gamma_{\mu}G_{D}^{*}(k)\sigma_{\nu\rho}\left(k-q\right)_{\rho}G_{D}^{*}(q)\right] \nonumber \\
& = & -\frac{f_{V}g_{V}^{2}M^{*}}{32\pi^{4}M}
\int_{0}^{k_{F}}\frac{|\mathbf{k}|^{2}d|\mathbf{k}|}{E^{*}(k)}
\int_{0}^{k_{F}}\frac{|\mathbf{q}|^{2}d|\mathbf{q}|}{E^{*}(q)} \int_{-1}^{1}d(\cos\theta)
\nonumber \\[7pt]
& & \times \left(\left[2E^{*}(k)E^{*}(q)-2|\mathbf{k}||\mathbf{q}|
\cos\theta-2{M^{*}}^{2}+M_{S}^{2}\right]\right. \nonumber \\
& & \times \left\{8\left(\left[2E^{*}(k)E^{*}(q)-2|\mathbf{k}||\mathbf{q}|
\cos\theta-2{M^{*}}^{2}+M_{V}^{2}+\frac{\zeta_{0}}{2}g_{V}^{2}V_{0}^{2}\right]\right.\right. \nonumber \\
& & \times \left. \left[M_{V}^{2}+\frac{\zeta_{0}}{6}g_{V}^{2}V_{0}^{2}\right]
-\frac{\zeta_{0}}{3}g_{V}^{2}V_{0}^{2}\left[E^{*}(k)-E^{*}(q)\right]^{2}\right) \nonumber \\
& & \times \left[E^{*}(k)E^{*}(q)-|\mathbf{k}||\mathbf{q}|
\cos\theta-{M^{*}}^{2}\right] \nonumber \\
& & +\frac{\zeta_{0}}{3}g_{V}^{2}V_{0}^{2}\left[M_{V}^{2}+\frac{\zeta_{0}}{6}g_{V}^{2}V_{0}^{2}\right] \nonumber \\
& & \left. \times \left(\left[2E^{*}(k)E^{*}(q)-2|\mathbf{k}||\mathbf{q}|\cos\theta-2{M^{*}}^{2}\right]
+\left[E^{*}(k)-E^{*}(q)\right]^{2}\right)\right\} \nonumber \\
& & -\eta^{2}V_{0}^{2}\left[8\left\{\left[E^{*}(k)-E^{*}(q)\right]^{2}
-\left(M_{V}^{2}+\frac{\zeta_{0}}{6}g_{V}^{2}V_{0}^{2}\right)\right\}\right. \nonumber \\
& & \times \left[E^{*}(k)E^{*}(q)-|\mathbf{k}||\mathbf{q}|\cos\theta-{M^{*}}^{2}\right] \nonumber \\
& & +\left(M_{V}^{2}+\frac{\zeta_{0}}{6}g_{V}^{2}V_{0}^{2}\right)
\left\{2\left[E^{*}(k)E^{*}(q)-|\mathbf{k}||\mathbf{q}|\cos\theta-{M^{*}}^{2}\right]\right. \nonumber \\
& & \left.\left.\left. +\left[E^{*}(k)-E^{*}(q)\right]^{2}\right\}\right]\right)
/ B \ ,
\end{eqnarray}

\noindent which were reduced in the same manner as in the previous discussion. Here $B$ is defined
in Eq.\ (\ref{eqn:den}).

\end{document}